\def\webDOI{https://doi.org/}

\documentclass[review,1p,times]{amsart}

\usepackage{epsfig}

\usepackage{amssymb}
\usepackage{amsthm}
\usepackage{amsmath,stmaryrd}
\usepackage{color}
\usepackage[pdftex, pdfborderstyle={/S/U/W 1}]{hyperref}
\usepackage{url}
\usepackage{cite}
\usepackage{nccmath}
\usepackage{cancel}

\usepackage{lineno}

\newcommand{\eref}[1]{Eq.~\eqref{#1}} 
\newcommand{\erefs}[2]{Eqs.~\eqref{#1}--\eqref{#2}} 
\newcommand{\sref}[1]{Sect.~\ref{#1}} 
\newcommand{\PBS}[1]{\let\temp=\\#1\let\\=\temp}

\newcommand{\aref}[1]{Appendix \ref{#1}}

\newcommand{\ci}{\mathrm{i}}

\newcommand{\itr}{{\sf T}}
\newcommand{\xgj}{x}
\newcommand{\kj}{k}
\newcommand{\xj}{x}

\newcommand{\speci}{w}

\newcommand{\speciv}{{\bf\speci}}
\newcommand{\eigl}{\omega}
\newcommand{\jeig}{\alpha}
\newcommand{\keig}{\beta}

\newcommand{\proj}{\boldsymbol{\Pi}}

\newcommand{\Go}{\mathrm{O}}
\newcommand{\po}{\mathrm{o}}

\newcommand{\vyj}{y}
\newcommand{\yv}{{\boldsymbol\vyj}}
\newcommand{\xg}{{\boldsymbol\xgj}}
\newcommand{\kv}{{\boldsymbol\kj}}
\newcommand{\hkv}{{\hat\kv}}
\newcommand{\xv}{{\boldsymbol\xj}}
\newcommand{\uv}{{\boldsymbol u}}
\newcommand{\vv}{{\boldsymbol v}}
\newcommand{\vpj}{p}
\newcommand{\vqj}{q}
\newcommand{\vp}{{\boldsymbol\vpj}}
\newcommand{\vq}{{\boldsymbol\vqj}}
\newcommand{\hvp}{\hat{\vp}}

\newcommand{\bzero}{{\bf 0}}		    
\newcommand{\eiga}{{\boldsymbol a}}
\newcommand{\eigb}{{\boldsymbol b}}
\newcommand{\eigc}{{\boldsymbol c}}			
\newcommand{\kgj}{k}
\newcommand{\kg}{{\boldsymbol\kgj}}
\newcommand{\hkg}{{\hat\kg}}
\newcommand{\pg}{{\boldsymbol p}}
\newcommand{\hpg}{{\hat\pg}}
\newcommand{\ev}{\hat{\boldsymbol e}}

\newcommand{\wg}{\omega}					

\newcommand{\cel}{c}
\newcommand{\Cel}{C}

\newcommand{\domain}{{\mathcal O}}			

\newcommand{\Id}{{\boldsymbol I}}

\newcommand{\iD}{\mathrm{D}}
\newcommand{\dd}{{\mathrm d}}				

\newcommand{\bnabla}{{\boldsymbol\nabla}}		
\newcommand{\grad}{{\boldsymbol\nabla}}
\newcommand{\Dx}{{\bf D}}				
\newcommand{\Dxx}{\Dx_\xg}
\newcommand{\Dxt}{\Dx_\xt}

\newcommand{\Dt}{\iD_t}	
\newcommand{\TF}[1]{\widehat{#1}}				
\newcommand{\iexp}{\operatorname{e}}			
\newcommand{\trace}{\operatorname{Tr}}		

\newcommand{\cjg}[1]{\overline{#1}}
\newcommand{\adj}[1]{{#1}^*}
\newcommand{\norm}[1]{|#1|}

\newcommand{\abs}[1]{\left|#1\right|}
\newcommand{\esp}[1]{{\mathbb E}\left\{#1\right\}}
\newcommand{\scal}[1]{\left\langle#1\right\rangle}
\newcommand{\scald}[1]{\left(#1\right)_{L^2}}
\newcommand{\Real}{\Re\mathfrak{e}}

\newcommand{\Rset}{\mathbb{R}}			
\newcommand{\Cset}{\mathbb{C}}
\newcommand{\Nset}{\mathbb{N}}

\newcommand{\diag}{\operatorname{diag}}              

\newcommand{\vl}{{\boldsymbol\ell}}
\newcommand{\vL}{{\boldsymbol L}}

\newcommand{\demi}{\frac{1}{2}}	      

\newcommand{\half}{\text{0.5}}

\newcommand{\obsj}{L}
\newcommand{\obsv}{\boldsymbol{\obsj}}
\newcommand{\av}{{\boldsymbol a}}

\newcommand{\coroij}{R}
\newcommand{\coro}{{\boldsymbol\coroij}}        

\newcommand{\indL}{1}

\newcommand{\iref}{0}
\newcommand{\escale}{\varepsilon}

\newcommand{\Wigner}{W}				
\newcommand{\Wignerv}{{\boldsymbol\Wigner}}
\newcommand{\Wignere}{\Wigner_\escale}
\newcommand{\Wignerve}{\Wignerv_\escale}

\newcommand{\uve}{\uv_\escale}
\newcommand{\vve}{\vv_\escale}

\newcommand{\cref}{\cel_\iref}
\newcommand{\Cref}{\Cel_\iref}

\newcommand{\xt}{{\boldsymbol s}}
\newcommand{\xte}{\frac{\xt}{\escale}}
\newcommand{\yt}{{\boldsymbol\tau}}
\newcommand{\kw}{{\boldsymbol\xi}}
\newcommand{\pw}{{\boldsymbol\eta}}

\newcommand{\sequence}[1]{(#1)}

\newcommand{\dscatij}{\sigma}
\newcommand{\dscat}{{\boldsymbol\dscatij}}
\newcommand{\tscati}{\Sigma}
\newcommand{\tscat}{{\boldsymbol\tscati}}

\newcommand{\lcor}{\ell_c}

\newcommand{\Bj}{B}
\newcommand{\Bv}{{\boldsymbol\Bj}}
\newcommand{\Dj}{D}
\newcommand{\Dv}{{\boldsymbol\Dj}}
\newcommand{\Ej}{E}
\newcommand{\Ev}{{\boldsymbol\Ej}}
\newcommand{\Jj}{J}
\newcommand{\Jv}{{\boldsymbol\Jj}}
\newcommand{\Hj}{H}
\newcommand{\Hv}{{\boldsymbol\Hj}}

\newcommand{\EHpv}{\begin{pmatrix} \Ev\\ \Hv\end{pmatrix}}
\newcommand{\permitj}{\epsilon}
\newcommand{\permeaj}{\mu}
\newcommand{\magnetoj}{\xi}
\newcommand{\permitv}{{\boldsymbol\permitj}}
\newcommand{\permeav}{{\boldsymbol\permeaj}}
\newcommand{\magnetov}{{\boldsymbol\magnetoj}}
\newcommand{\impj}{Z}
\newcommand{\Kv}{{\boldsymbol K}}
\newcommand{\Kper}{{\boldsymbol V}}
\newcommand{\kper}{{\bf s}}
\newcommand{\Mv}{{\boldsymbol M}}
\newcommand{\Nv}{{\boldsymbol N}}
\newcommand{\Pv}{{\boldsymbol P}}
\newcommand{\Rv}{{\boldsymbol R}}
\newcommand{\nv}{{\boldsymbol n}}
\newcommand{\Omv}{{\boldsymbol\Omega}}
\newcommand{\deng}{{\mathcal E}}

\newcommand{\Flowj}{{\mathcal F}}
\newcommand{\Flow}{\boldsymbol\Flowj}
\newcommand{\Maxwell}{\boldsymbol M}
\newcommand{\vT}{\boldsymbol T}
\newcommand{\vX}{\boldsymbol X}

\newcommand{\alert}[1]{\textcolor{black}{#1}}

\newtheorem{remark}{Remark}
\newtheorem{example}{Example}

\begin{document}




\title[Wigner measures in bianisotropic media]{Wigner measures of electromagnetic waves in heterogeneous bianisotropic media}


\author[J.-L. Akian]{Jean-Luc Akian}
\address[J.-L. Akian]{ONERA--The French Aerospace Lab, France}
\email{jean-luc.akian@onera.fr}

\author[\'E. Savin]{\'Eric Savin}
\address[\'E. Savin]{ONERA--The French Aerospace Lab, France}
\thanks{Corresponding author: \'E. Savin, ONERA--The French Aerospace Lab, 6 chemin de la Vauve aux Granges, FR-91123 Palaiseau cedex, France (eric.savin@onera.fr).}
\email{eric.savin@onera.fr}

\begin{abstract}
We study the propagation of high-frequency electromagnetic waves in randomly heterogeneous bianisotropic media with dissipative properties. For that purpose we consider randomly fluctuating optical responses of such media with correlation lengths comparable to the typical wavelength of the waves. Although the fluctuations are weak, they induce multiple scattering over long propagation times and/or distances such that the waves end up travelling in many different directions with mixed polarizations. We derive the dispersion and evolution properties of the Wigner measure of the electromagnetic fields, which describes their angularly-resolved energy density in this propagation regime. The analysis starts from Maxwell's equations with general constitutive equations. We first ignore the random fluctuations of the optical response and obtain {\em uncoupled} transport equations for the components of the Wigner measure on the different propagation modes (polarizations). Then we use a multi-scale expansion of the Wigner mesure to obtain the radiative transfer equations satisfied by these components when the fluctuations are no longer ignored. The radiative transfer equations are {\em coupled} through their collisional parts, which account for the scattering of waves by the random fluctuations and their possible changes in polarization. The collisional kernels describing these processes depend on the power and cross-power spectral densities of the fluctuations at the wavelength scale. The overall derivation is based on the interpretation of Wigner transforms and Wigner measures in terms of semiclassical pseudo-differential operators in their standard quantization. 
\end{abstract}


\date{\today}


\maketitle

\section{Introduction}


The analysis of electromagnetic wave propagation in complex, heterogeneous media has relevance to a wide range of engineering applications, including radar imaging and remote sensing \cite{SKO08}, optical imaging \cite{CAR21}, atmospheric physics \cite{ISH78}, or communications \cite{GOL05}, among others. As the complexity of the propagation media can not be known exactly but only assessed in a statistical sense, we model their inhomogeneities (or imperfections) as random fields and thus consider Maxwell's equations with randomly fluctuating electromagnetic parameters. This approach pertains to waveguides filled with dielectric materials having numerous imperfections, atmospheres and clouds, colloidal suspensions, biological tissues, or plasmas, which is the type of media we have more particularly in mind here.

Asymptotic methods can be used to characterize both quantitatively and qualitatively the wave fields when some explicit scales can be identified and separated to describe the corresponding propagation regimes \cite{AKK07,ROS99,SHE06}. Recent mathematical works on electromagnetic waves in random media have considered layered half-spaces \cite{GAR15}, waveguides \cite{ALO15}, or beam propagation in open environments \cite{BOR16}, for example. In these works a preferred direction of propagation can be identified, and the electromagnetic field can be decomposed into plane wave components transverse to that direction. Their amplitudes are random fields of which evolution is driven by the random fluctuations of the dielectric parameters and can be studied in different asymptotic regimes. For example, the cumulative scattering effects induced by weak inhomogeneities at large distances of propagation in a waveguide or a beam are typically manifested in the exponential decay of the average fields, the enhancement of their random fluctuations, and depolarization.

We rather investigate in this paper the situation where no preferred direction of propagation can be identified beforehand as waves travel in open random media. The wavelength and the correlation length of the random fluctuations of the electromagnetic parameters are small and comparable, while the propagation distance and/or time lapse are large. The amplitude of the random fluctuations is also weak. This high-frequency scaling is of interest if the objective is for instance to probe some inhomogeneities from remote sensors. In this setting, waves are multiply scattered by the inhomogeneities as they travel in all directions, and their polarization and phase get randomized. It is then known that the energy density is the relevant quantity to focus on, and that it can be evaluated from the Wigner transform of the wave field and its non-negative high-frequency limit measure--the Wigner measure \cite{GER97,LIO93,MAR02,ZWO12}. The evolution of the Wigner measure is described by radiative transfer equations \cite{CHA60,ISH78,MIS14,TSA00} with linear collisional operators of which kernels depend on the correlation structure of the inhomogeneities \cite{BAL10,RYZ96}. Actually, the sole information needed to account for the fluctuations of the electromagnetic parameters is their power spectral densities, \emph{i.e.} the spatial Fourier transforms of their auto-correlation functions. These radiative transfer equations for electromagnetic vector waves were already derived in \cite{PAP75,RYZ96} for random fluctuations in isotropic media, and subsequently in \cite{FAN07b,FAN07c} for random fluctuations in bianisotropic media with homogeneous mean background parameters and no dissipation. Our aim is to extend these results to more general constitutive equations, for heterogeneous materials with dispersive and dissipative properties.

Radiative transfer equations have a geometrical interpretation \cite{LES15} notably in terms of bicharacteristic rays that is, rays in phase space, which parallels classical asymptotic ray methods \cite{BAB91,KRA90}. The latter are applicable to synthetic aperture radar (SAR) simulation for example \cite{AUE10}, including the investigation of multipath and vibration signatures in radar imaging and remote sensing or the computation of antenna transfer functions in heterogeneous media whenever the approximations of geometrical optics hold. Ray tracing solvers are needed in this respect \cite{GLA89}. One of the possible applications of this research concerns the development of ray tracing algorithms in phase space for complex media such as plasmas \cite{TRA14}. Ray tracing has been intensively developed in the last decade in the context of computer graphics \cite{PAR10,HAI19}, notably, and one also expects that the application programming interfaces that have been implemented in the video game industry, among others, could be used in the context of computational electromagnetics alike.

The rest of the paper is organized as follows. In \sref{sec:HFwaves} we introduce the basic physical framework and notations used throughout, and we summarize our main results. The relevance of considering a Wigner transform and its non-negative limit measure for the analysis of high-frequency wave propagation phenomena is emphasized. We first consider the transport regime in the absence of fluctuations of the optical response that models instantaneous effects, and then the radiative transfer regime when weak random fluctuations are taken into account. The detailed derivation of the transport equations is outlined in \sref{sec:Transport}, and the detailed derivation of the radiative transfer equations is outlined in \sref{sec:RTE}. We use pseudo-differential calculus and the interpretation of Wigner transforms in terms of semiclassical pseudo-differential operators, choosing the same quantization for both of them. We argue that this choice clarifies the analyses as compared to other approaches where the Wigner transforms and the operators have different quantizations. A summary and outlook are finally drawn in \sref{sec:CL}.

\section{Maxwell's equations in bianisotropic media and main results}\label{sec:HFwaves}

\subsection{Physical setting}

We wish to analyze high-frequency electromagnetic wave propagation phenomena in general heterogeneous, bianisotropic dielectric media. The Amp\`ere and Faraday laws read:
\begin{equation}\label{eq:AmpereFaraday}
\frac{\partial\Dv}{\partial t}=\bnabla\times\Hv-\Jv\,,\quad\frac{\partial\Bv}{\partial t}=-\bnabla\times\Ev\,,
\end{equation}
where $\Bv$ is the magnetic flux density, $\Dv$ is the electric flux density, or electric displacement field, $\Ev$ is the electric field, $\Hv$ is the magnetic field, and $\Jv$ is the density of electric current. Taking the divergence of \eref{eq:AmpereFaraday} the fluxes are subjected to Gauss laws:
\begin{equation}\label{eq:Gauss}
\bnabla\cdot\Dv=\rho\,,\quad\bnabla\cdot\Bv=0\,,
\end{equation}
for $\rho$ being the density of electric charge, which is then related to the current by the equation of continuity of charge:
\begin{displaymath}
\frac{\partial\rho}{\partial t}+\bnabla\cdot\Jv=0\,.
\end{displaymath}
Here we consider open media in $\Rset^3$ with no external charge ($\rho=0$) and no external current ($\Jv=\bzero$). Besides, Gauss laws hold for all times provided that they hold at some initial time. The Amp\`ere and Faraday laws are supplemented with constitutive equations for general linear dispersive media with the properties of causality, time-invariance, and locality (see \emph{e.g.} \cite{CAS17,IOA12,KAR92}), which read:
\begin{equation}\label{eq:constitutive}
\begin{split}
\Dv(\xv,t) &=\permitv_\iref(\xv)\Ev(\xv,t)+(\permitv_d\star\Ev)(\xv,t) +  \magnetov_\iref (\xv)\Hv(\xv,t) + (\magnetov_d\star\Hv)(\xv,t)\,, \\
\Bv(\xv,t) &={\boldsymbol\zeta}_\iref(\xv)\Ev(\xv,t)+({\boldsymbol\zeta}_d\star\Ev)(\xv,t) +  \permeav_\iref (\xv)\Hv(\xv,t) + (\permeav_d\star\Hv)(\xv,t)\,,
\end{split}
\end{equation}
where $\permitv_\iref(\xv)$ is the spatially variable permittivity tensor, $\permeav_\iref(\xv)$ is the spatially variable permeability tensor, and $\magnetov_\iref(\xv)$ and ${\boldsymbol\zeta}_\iref(\xv)$ are spatially variable magnetoelectric tensors. Also $\star$ stands for the time convolution product:
\begin{displaymath}
(\permitv_d\star\Ev)(\xv,t)=\int_0^t\permitv_d(\xv,\tau)\Ev(\xv,t-\tau)\dd\tau
\end{displaymath}
with similar expressions for the other terms. \alert{The upper bound $t$ is because one may wish to consider initial data at $t=0$ in the subsequent analyses; see \sref{sec:WKB} or Example~\ref{example2} below}. In \eref{eq:constitutive} $\Pv=\smash{\permitv_d\star\Ev}$ stands for the polarization field and $\Mv=\smash{\permeav_\iref^{-1}(\permeav_d\star\Hv)}$ stands for the magnetization field, but these constitutive equations also account for coupled magnetoelectric effects through the magnetoelectric tensors. Here we consider the subclass of bianisotropic materials for which the following symmetry relationships hold: $\smash{{\boldsymbol\zeta}_\iref}=\smash{\adj{\magnetov}_\iref}$, \emph{i.e.} the conjugate transpose of $\smash{\magnetov_\iref}$; and $\smash{\permitv_\iref}=\smash{\adj{\permitv}_\iref}$, $\smash{\permeav_\iref}=\smash{\adj{\permeav}_\iref}$. These symmetry conditions reduce the number of independent constitutive parameters in $\permitv_\iref$, $\smash{\permeav_\iref}$, and $\smash{\magnetov_\iref}$ to 21.

Now plugging \eref{eq:constitutive} into \eref{eq:AmpereFaraday}, we arrive at Maxwell's $6\times 6$ system for the electric field $\Ev$ and magnetic field $\Hv$:
\begin{equation}\label{eq:DampedMaxwell}
\frac{\partial}{\partial t}\left(\begin{bmatrix} \permitv_\iref & \magnetov_\iref \\ \adj{\magnetov}_\iref & \permeav_\iref \end{bmatrix} + \begin{bmatrix} \permitv_d & \magnetov_d \\ {\boldsymbol\zeta}_d & \permeav_d \end{bmatrix}\star\right)\EHpv + \grad\times\begin{bmatrix} \bzero & -\Id \\ \Id & \bzero \end{bmatrix}\EHpv = \bzero\,,
\end{equation}
where $\Id$ is the $3\times 3$ identity matrix. We introduce the following matrices of electromagnetic tensors:
\begin{equation}
\Kv_\iref(\xv)=\begin{bmatrix} \permitv_\iref(\xv) & \magnetov_\iref(\xv) \\ \adj{\magnetov}_\iref(\xv) & \permeav_\iref(\xv) \end{bmatrix}\,,\quad\Kv_d(\xv,t)=\begin{bmatrix} \permitv_d(\xv,t) & \magnetov_d(\xv,t) \\ {\boldsymbol\zeta}_d(\xv,t) & \permeav_d(\xv,t)\end{bmatrix}\,.
\end{equation}
$\smash{\Kv_\iref}$ is called the optical response, or (generalized) susceptibility tensor, and models instantaneous effects, while $\smash{\Kv_d}$ is called the (generalized) susceptibility kernel, and models memory as well as dissipation effects \cite{IOA12,KAR92}. We assume throughout the paper that $\smash{\Kv_\iref}$, which is Hermitian, is also positive definite, as in \cite{FAN07b,FAN07c}. A negative definite optical response gives rise to a negative index of refraction, as in metamaterials \cite{VES68}.

\begin{example}
An isotropic, lossless medium is $\smash{\Kv_\iref(\xv)}=\smash{\diag(\permitj_\iref(\xv)\Id, \permeaj_\iref(\xv)\Id)}$ and $\smash{\Kv_d}=\bzero$. A chiral medium is a reciprocal bi-isotropic medium with $\smash{\permitv_\iref}=\smash{\permitj_\iref}\Id$, $\smash{\permeav_\iref}=\smash{\permeaj_\iref}\Id$, and a purely imaginary magnetoelectric tensor $\smash{\magnetov_\iref}=\ci\chi\Id$, where $\smash{\permitj_\iref}$ and $\smash{\permeaj_\iref}$ are the permittivity and permeability constants, and $\chi\in\Rset$ is the magnetoelectric constant. An example of a dissipative, local homogeneous material (in the terminology of \cite{CAS17}) is the Lorentz model with damping in \cite{RAM10} with:
\begin{equation}\label{eq:Lorentz}
\Kv_\iref=\begin{bmatrix} \permitj_\iref\Id & \bzero \\ \bzero & \permeaj_\iref\Id \end{bmatrix}\,,\quad\TF{\Kv}_d(\wg)=\begin{bmatrix} \TF{\permitj}_d(\wg)\Id & \bzero \\ \bzero & \bzero \end{bmatrix}\,,
\end{equation}
where $\TF{\permitj}_d(\wg)=\smash{\frac{\permitj_\iref\wg_p^2}{-\wg^2+\ci\wg\Gamma+\wg_0^2}}$ is the susceptibility (in frequency domain), $\wg_p$ is the plasma frequency such that $\smash{\wg_p^2}=\smash{\frac{Ne^2}{m\permitj_\iref}}$, $\wg_0$ is a characteristic frequency for the motion of an electron with mass $m$ and charge $e$, $N$ is the electron density, and $\Gamma$ is a damping loss rate. The Fourier transform in time domain is $\smash{\TF{\Kv}_d}(\wg)=\smash{\int_\Rset\iexp^{-\ci\wg t}\Kv_d(t)\dd t}$. Drude material is $\Gamma=0$ and $\smash{\wg_0=0}$. Other examples are described in {\em e.g.} \cite{CAS17,FAN07b,FAN07c,KAR92}.
\end{example}

We then introduce the scalar product $\scal{\uv,\vv}=\adj{(\Kv_\iref\uv)}\vv=\adj{\uv}\Kv_\iref\vv$ (for $\smash{\adj{\Kv}_\iref}=\smash{\Kv_\iref}$ where $\smash{\adj{\Kv}_\iref}$ stands for the conjugate transpose of $\smash{\Kv_\iref}$), and the electromagnetic energy density $\deng(\xv,t) $ and the energy flow density (Poynting vector) $\Flow(\xv,t)$:
\begin{equation}\label{eq:energy}
\begin{split}
\deng(\xv,t) &=\demi\scal{\EHpv,\EHpv} \\
&=\demi\adj{\Ev(\xv,t)}\permitv_\iref(\xv)\Ev(\xv,t) + \demi\adj{\Hv(\xv,t)}\permeav_\iref(\xv)\Hv(\xv,t) \\
&\quad\quad+\Real\{\adj{\Ev(\xv,t)}\magnetov_\iref(\xv)\Hv(\xv,t)\} \,, \\
\Flow(\xv,t) &=\cjg{\Ev(\xv,t)}\times\Hv(\xv,t)\,,
\end{split}
\end{equation}
\alert{for $\Re\mathfrak{e}$ standing for the real part}, such that in the undamped case $\smash{\Kv_d(\xv,t)}\equiv\bzero$:
\begin{displaymath}
\frac{\partial\deng}{\partial t}+\grad\cdot\Flow=0\,,\quad\frac{\dd }{\dd t}\int\deng(\xv,t)\dd\xv=0\,.
\end{displaymath}
The conservation law above establishes how the energy density is spread in space, but it does not describe how it propagates. Our main objective in this paper is to describe how this quantity evolves both in space and direction for high-frequency waves induced by, say, strongly oscillating initial conditions, and for a randomly fluctuating optical response and a non-vanishing susceptibility kernel $\smash{\Kv_d}$. In the next section we show how this goal can be achieved using Wigner transforms of the wave fields and their high-frequency limits. We illustrate it with the wave equation with constant speed, before we turn to Maxwell's system \eqref{eq:DampedMaxwell} and state our main results in \sref{sec:MainResults}. 

\subsection{High-frequency limit and Wigner transform}\label{sec:WKB}

Maxwell's equations (\ref{eq:DampedMaxwell}) are subjected to highly oscillating initial data for the electromagnetic field $\uv=(\Ev,\Hv)$ at $t=0$. Consider for example the WKB form \cite{GER97,FIL03,SPA03}:
\begin{equation}\label{eq:CI}
\uv(\xv,0)=\uve^\text{WKB}(\xv)=\Re\mathfrak{e}\{\uv_I(\xv)\iexp^{\frac{\ci}{\escale}S_I(\xv)}\}\,,
\end{equation}
where $\smash{\uv_I}$ is a square integrable function on $\Rset^3$; and $\smash{S_I}$ is an integrable scalar function on $\Rset^3$ (at least locally), as well as its derivative $\smash{\bnabla_\xv S_I}$. The small parameter $\escale$ represents the typical wavelength of oscillations of these data. These oscillations are inherited by the actual electromagnetic field $\uv(\xv,t)$ satisfying (\ref{eq:DampedMaxwell}) at all times, which prevents it from converging nicely as the wavelength $\escale$ gets smaller and smaller in the high-frequency limit. A way to tackle this shortcoming is to consider quadratic observables of $\uv$ instead, as emphasized in \emph{e.g.} \cite{AKI12,BAL05,BAL10,BAY14,GOM12,GUO99,POW05,RYZ96} for different types of waves. In particular, the Wigner transform \cite{GER97,LIO93,MAR02,ZWO12} of the solutions of \eref{eq:DampedMaxwell} shall be considered since its evolution in the limit $\escale\to 0$ can be derived explicitly. It provides a phase space description of how the associated energy density propagates in this very limit. The advantages of this approach over more classical methods such as the WKB method, which is suggested by the type of data (\ref{eq:CI}), are discussed in \cite{FIL03,SPA03} for example. Notably, the Wigner transform needs much lower regularity assumptions on $\smash{\uv_I}$ and $\smash{S_I}$ than in the WKB method, and evades the singularities (focal points or caustics) that can develop in finite times with the latter method.

Let $\vartheta\in[0,1]$. For \alert{tempered} distributions $\uv,\vv$ defined on $\Rset^m$, the Wigner transform is:
\begin{equation}\label{eq:TWigner}
\Wignerve^\vartheta[\uv,\vv](\xv,\kv)=\frac{1}{(2\pi)^m}\int_{\Rset^m}\iexp^{\ci\kv\cdot\yv}\uv\left(\xv-\escale(1-\vartheta)\yv\right)\adj{\vv}\left(\xv+\escale\vartheta\yv\right)\,\dd\yv\,.
\end{equation}
We denote by $\Wignerve^\vartheta[\uv]:=\Wignerve^\vartheta[\uv,\uv]$ the self Wigner transform of $\uv$. Now let $\sequence{\uve}$ be a bounded sequence in $\smash{L^2(\Rset^m)}$, the set of square integrable functions defined on $\Rset^m$. Then it can be shown that, up to extracting a subsequence if need be, $\Wignerve^\vartheta[\uve]$ has a weak limit (in the sense of \alert{tempered} distribution) as $\escale\to 0$ which is independent of the choice of $\vartheta$ in \eref{eq:TWigner}. Also if $\Wignerv[\uve]$ is such a limit, it is a non-negative, Hermitian matrix-valued measure. These results are detailed in \emph{e.g.} \cite{GER97,ZWO12}. Here the notation of \cite[p.~330]{GER97} is used for the limit $\Wignerv$ (independent of $\escale$) of the family $\smash{\sequence{\uve}}$ (dependent of $\escale$) but clearly the measure $\Wignerv[\uve]$ is independent of $\escale$. The scalar Wigner measure is the trace of the latter $\Wigner[\uve]=\trace\Wignerv[\uve]$ and can be directly related to the energy density of the sequence $\sequence{\uve}$ in the high-frequency limit. Taking the Wigner transform of the sequence of WKB data (\ref{eq:CI}) for example, one obtains:
\begin{displaymath}
\begin{split}
\Wignerv[\uve^\text{WKB}](\xv,\kv) &=\uv_I(\xv)\adj{\uv}_I(\xv)\delta(\kv-\bnabla_\xv S_I(\xv))\,, \\
\Wigner[\uve^\text{WKB}](\xv,\kv) &=\norm{\uv_I(\xv)}^2\delta(\kv-\bnabla_\xv S_I(\xv))\,.
\end{split}
\end{displaymath}
If these data are propagated by the wave equation in $\Rset^3$ with constant speed $\cel$:
\begin{displaymath}
\begin{split}
\frac{1}{\cel^2}\frac{\partial^2\uve}{\partial t^2}-\Delta\uve=\bzero\,, &\quad \xv\in\Rset^3\,,\;t>0\,, \\
\uve(\xv,0)=\bzero\,,\;\frac{\partial\uve}{\partial t}(\xv,0)=\uve^\text{WKB}(\xv)\,, &\quad \xv\in\Rset^3\,,
\end{split}
\end{displaymath}
then for $t>0$ Kirchhoff's formula yields \cite[\S 2.4]{EVA10}:
\begin{displaymath}
\uve(\xv,t)=\int_{S^2}t\uve^\text{WKB}(\xv-\cel\hpg t)\frac{\dd\Omega(\hpg)}{4\pi}\,,
\end{displaymath}
where $S^2$ is the unit sphere of $\Rset^3$ centered at $\bzero$, and $\dd\Omega$ is its surface differential element. The scalar Wigner measure of $\uve(\cdot,t)$ reads \cite[\S 3.3]{GER97}:
\begin{displaymath}
\Wigner[\uve](\xv,\kv,t)=\demi\sum_{\jeig=\pm}\norm{\smash{\uv_I(\xv+\jeig\cel\hkv t)}}^2\delta(\kv-\bnabla_\xv S_I(\xv+\jeig\cel\hkv t))\,,
\end{displaymath}
with the usual notation $\hkv=\smash{\frac{\kv}{\abs{\kv}}}$ for $\kv\neq\bzero$. It provides a characterization of the evolution of the associated energy density both in space and direction. Because of the assumed low regularity of $\smash{\uv_I}$ and $\smash{S_I}$, a classical WKB method could not be used with these data. Note that more general data can be considered as well as shown in \cite{GER97}.

\subsection{Main results}\label{sec:MainResults}

We derive the Wigner measure $\Wignerv$ of the high-frequency electromagnetic field $\uv$ which obeys Maxwell's $6\times 6$ system \eqref{eq:DampedMaxwell} when the optical response $\smash{\Kv_\iref}$ exhibits in addition random fluctuations. That is:
\begin{equation}\label{eq:RandomMaxwell}
\frac{\partial}{\partial t}\left(\Kv(\xv)\uv(\xv,t)+\int_0^t\Kv_d(\xv,\tau)\uv(\xv,t-\tau)\dd\tau\right)+\Maxwell(\bnabla_\xv)\uv(\xv,t)=\bzero
\end{equation}
for $\xv,t\in\domain\times\Rset_+^*$ in some open domain $\domain\subseteq\smash{\Rset^3}$ and for highly oscillating initial conditions at $t=0$ characterized by the small scale $\escale\ll 1$. Here:
\begin{equation}\label{eq:MaxwellO}
\Maxwell(\kv)=\begin{bmatrix} \bzero & -\Omv(\kv) \\ \Omv(\kv) & \bzero \end{bmatrix}
\end{equation}
is the symmetric Maxwell operator for $\Omv(\kv)$ being the skew-symmetric matrix such that $\Omv(\kv)\av=\kv\times\av$ for any vector $\av\in\smash{\Cset^3}$, \alert{and accordingly:
\begin{displaymath}
\Maxwell(\bnabla_\xv) = \begin{bmatrix} \bzero & -\bnabla_\xv\times \\ \bnabla_\xv\times & \bzero \end{bmatrix}\,.
\end{displaymath}}
Also the actual optical response $\Kv(\xv)$ reads:
\begin{displaymath}
\Kv(\xv) = \Kv_\iref(\xv)\left[\Id+\sigma\Kper\left(\frac{\xv}{\lcor}\right)\right]\,,
\end{displaymath}
where the dimensionless random matrix $\Kper$ represents random fluctuations of the optical response $\smash{\Kv_\iref}$ of the background medium, and is such that $\smash{\Kv_\iref}\Kper=\adj{\Kper}\smash{\Kv_\iref}$ to preserve the Hermiticity of the actual optical response $\Kv$. It is assumed that $(\Kper(\yv),\,\yv\in\smash{\Rset^3})$ is a second-order, mean-square homogeneous (spatially stationary) random field with zero mean $\esp{\Kper(\yv)}=\bzero$ and integrable fourth-order autocorrelation tensor $\coro$ such that:
\begin{equation}\label{eq:covariance}
\coro(\yv-\yv'):=\esp{\Kper(\yv)\otimes\Kper(\yv')}=\int_{\Rset^3}\iexp^{\ci\kv\cdot(\yv-\yv')}\TF{\coro}(\kv)\dd\kv\,,
\end{equation}
where $\esp{\cdot}$ stands for the mathematical expectation, or average. The second equality above stems from Bochner's theorem and the fact that $\yv\mapsto\coro(\yv)$ is positive semi-definite. The power spectral density of that random field is the matrix:
\begin{equation}\label{eq:Bochner}
\TF{\coro}(\kv)=\frac{1}{(2\pi)^3}\int_{\Rset^3}\iexp^{-\ci\kv\cdot\yv}\coro(\yv)\dd\yv
\end{equation}
which is positive. The autocorrelation tensor is normalized such that $\smash{\int_{\Rset^3}\coro(\yv)\dd\yv}=\Go(1)$ and $\coro(\bzero)=\Go(1)$. The length scale $\lcor$ is the correlation length and the dimensionless scalar $\sigma\geq 0$ quantifies the amplitude of the fluctuations. These parameters are of the same order for all correlations of the fluctuations in the proposed model.

At first, the random fluctuations of the optical response are disregarded, setting $\sigma=0$. \alert{We state our main results formally in the open medium $\domain$ but emphasize here that they can be made rigorous by the method outlined in Remark~\ref{remark1} below. Possible boundary effects are evaded by this approach, though. This is a difficult topic that could be partly addressed by the analysis carried on in \cite{AKI12}}. Let:
\begin{equation}\label{eq:PDOperators}
\begin{split}
\vL_\iref(\xv,\kv) &=\Kv_\iref^{-1}(\xv)\Maxwell(\kv)\,,\\
\vL_\indL(\xv,\wg) &=\ci\wg\Kv_\iref^{-1}(\xv)\TF{\Kv}_d(\xv,\wg)\,,
\end{split}
\end{equation}
where $\TF{\Kv}_d(\xv,\wg)$ is the Fourier transform of $\Kv_d(\xv,t)$ with respect to $t$. $\smash{\vL_\iref}$ being \alert{Hermitian} for the scalar product $\scal{\cdot}$, \alert{such that $\smash{\scal{\vL_\iref\uv,\vv}}=\smash{\scal{\uv,\vL_\iref\vv}}$}, let us introduce the eigen-expansion:
\begin{equation}\label{eq:spectralL0}
\vL_\iref=\sum_{\jeig}\eigl_\jeig\proj_\jeig
\end{equation}
where $\proj_\jeig$ is the projector on the $\jeig$-th eigen-subspace with associated real eigenvalue $\eigl_\jeig$, such that $\sum_{\jeig}\proj_\jeig=\Id$ and $\Kv_\iref\proj_\jeig=\adj{\proj}_\jeig\Kv_\iref$. The right eigenvectors $\smash{\eigb_\jeig}=\smash{(\eigb_\jeig^1,\dots\eigb_\jeig^A)}$ are such that $\smash{\scal{\eigb_\jeig,\eigb_\keig}}=\smash{\adj{(\Kv_\iref\eigb_\jeig)}\eigb_\keig}=\smash{\delta_{\jeig\keig}\Id_A}$, where $A$ is the order of multiplicity of the $\jeig$-th eigenvalue and $\Id_A$ is the $A\times A$ identity matrix; also $\sum A=6$. Letting $\eigc_\jeig=\Kv_\iref\eigb_\jeig$ be the left eigenvectors, one has therefore:
\begin{equation}
\begin{split}
\vL_0\eigb_\jeig &=\eigl_\jeig\eigb_\jeig\,, \\
\adj{\eigc}_\jeig\vL_0 &=\eigl_\jeig\adj{\eigc}_\jeig\,, \\
\proj_\jeig &=\eigb_\jeig\adj{\eigc}_\jeig\,.
\end{split}
\end{equation}
Then it is shown that:
\begin{equation}\label{eq:WignerProj}
\Wignerv=\sum_{\jeig}\delta(\wg+\eigl_\jeig)\Wignerv_\jeig\,,
\end{equation}
where $\Wignerv_\jeig=\proj_\jeig\Wignerv=\Wignerv\adj{\proj}_\jeig=\proj_\jeig\Wignerv\adj{\proj}_\jeig$ are $6\times 6$ matrix-valued measures for each mode of propagation $\jeig$.

Alternatively $\Wignerv$ may be written:
\begin{equation}
\Wignerv=\sum_{\jeig}\delta(\wg+\eigl_\jeig)\eigb_\jeig\speciv_\jeig\adj{\eigb}_\jeig\,,
\end{equation}
where $\speciv_\jeig=\adj{\eigc}_\jeig\Wignerv_\jeig\eigc_\jeig$ are $A\times A$ matrix-valued measures. In \sref{sec:Transport}, the latter are shown to satisfy the transport equations:
\begin{equation}\label{eq:Liouville}
\partial_t\speciv_\jeig+\{\eigl_\jeig,\speciv_\jeig\}  + \vl_\jeig\speciv_\jeig + \speciv_\jeig\adj{\vl}_\jeig + \nv_\jeig\speciv_\jeig - \speciv_\jeig\nv_\jeig = \bzero\,,
\end{equation}
where $\{\eigl_\jeig,\speciv_\jeig\}=\bnabla_\kv\eigl_\jeig\cdot\bnabla_\xv\speciv_\jeig-\bnabla_\xv\eigl_\jeig\cdot\bnabla_\kv\speciv_\jeig$ stands for the usual Poisson bracket, $\nv_\jeig$ is a $A \times A$ skew-symmetric matrix given below by \eref{eq:skew-symmatrix}, and $\smash{\vl_\jeig}=\smash{\adj{\eigc}_\jeig\vL_\indL\eigb_\jeig}$. Taking the trace of \eref{eq:Liouville} yields:
\begin{displaymath}
\partial_t\speci_\jeig+\{\eigl_\jeig,\speci_\jeig\}  + 2\trace(\vl_\jeig^s\speciv_\jeig) = 0\,,
\end{displaymath}
where $\speci_\jeig=\trace\speciv_\jeig=\trace(\Kv_\iref\Wignerv_\jeig)$, and $\smash{\vl_\jeig^s}=\smash{\demi(\vl_\jeig+\adj{\vl}_\jeig)}$ is the symmetric part of $\smash{\vl_\jeig}$. The high-frequency energy density and Poynting vector \eqref{eq:energy} are then:
\begin{equation}\label{eq:energyHF}
\begin{split}
\deng(\xv,t) &=\demi\int_{\Rset^3}\trace(\Kv_\iref(\xv)\Wignerv(\xv,\kv,t))\dd\kv=\demi\sum_\jeig\int_{\Rset^3}\speci_\jeig(\xv,\kv,t)\dd\kv\,, \\
\Flow(\xv,t) &=\demi\int_{\Rset^3}\trace(\bnabla_\kv\Maxwell(\kv)\Wignerv(\xv,\kv,t))\dd\kv=\demi\sum_\jeig\int_{\Rset^3}\speci_\jeig(\xv,\kv,t)\bnabla_\kv\eigl_\jeig\dd\kv\,.
\end{split}
\end{equation}
The evolution of the energy density and flow in phase space are thus described by $\sum_\jeig\speci_\jeig$ which can also select the different modes of propagation. We observe however from \eref{eq:PDOperators} with \eref{eq:MaxwellO}, that $\smash{\eigl_0}=0$ ($\jeig=\text{"0"}$) is always an eigenvalue of $\smash{\vL_\iref}$ with multiplicity $A=2$ and eigenvectors $\smash{(\hkg,\bzero)}$ and $\smash{(\bzero,\hkg)}$ for $\kv\neq\bzero$. This mode is non propagative, though.

Secondly, we analyze the influence of random fluctuations of the optical response $\smash{\Kv_\iref}$ of the background medium by setting $\sigma>0$. We consider the scattering regime where $\lcor$ is of the order of the typical wavelength $\lambda$, which is the scale of variation of the strongly oscillating initial data, \emph{i.e.} it is small with respect to the typical propagation distance $L$:
\begin{equation}\label{eq:lcor}
\frac{\lcor}{L}\approx\frac{\lambda}{L}=\escale\,.
\end{equation}
Besides, the amplitude of fluctuations is also small with the scaling:
\begin{equation}\label{eq:sigma}
\sigma^2\approx\frac{\lcor}{L}\,.
\end{equation}
For a large distance of propagation $L$ the electromagnetic waves are multiply scattered by the fluctuations of the background medium and their energy is spread over many directions of propagation. Considering these fluctuations, it is shown in \sref{sec:RTE} that the transport equations \eqref{eq:Liouville} are modified to the coupled radiative transfer equations:
\begin{multline}\label{eq:RTE}
\partial_t\speciv_\jeig+\{\eigl_\jeig,\speciv_\jeig\}+(\vl_\jeig+\tscat_\jeig)\speciv_\jeig +\speciv_\jeig\adj{(\vl_\jeig+\tscat_\jeig)} +\nv_\jeig\speciv_\jeig - \speciv_\jeig\nv_\jeig \\
=\sum_{\keig}\int_{\Rset^3}\delta(\eigl_\keig(\xv,\vp)-\eigl_\jeig(\xv,\kv))\dscat_{\jeig\keig}(\xv,\kv,\vp):\speciv_\keig(\xv,\vp,t)\dd\vp \,,
\end{multline}
where $\dscat_{\jeig\keig}$ is a linear operator called differential scattering cross-section, and $\tscat_\jeig$ is a $A\times A$ matrix called total scattering cross-section. They are explicitly given in terms of the power spectral density matrix \eqref{eq:Bochner} of the random fluctuations $\Kper$ (see \eref{eq:dscat} and \eref{eq:tscat}, respectively). The $A\times A\times B\times B$ tensor $\dscat_{\jeig\keig}(\xv,\kv,\vp)$ describes how, in the multiple scattering process, the $B\times B$ matrix $\speciv_\keig$ for the mode $\beta$ with multiplicity $B$ in the direction $\vp$ is locally converted to the $A\times A$ matrix $\speciv_\jeig$ for the mode $\jeig$ with multiplicity $A$ in the direction $\kv$:
\begin{displaymath}
[\dscat_{\jeig\keig}(\xv,\kv,\vp):\speciv_\keig(\xv,\vp,t)]_{aa'}=\sum_{1\leq b,b'\leq B}[\dscat_{\jeig\keig}(\xv,\kv,\vp)]_{aa'bb'}[\speciv_\keig(\xv,\vp,t)]_{bb'}
\end{displaymath}
for $1\leq a,a'\leq A$. The $A\times A$ matrix $\tscat_\jeig(\xv,\kv)$ describes how the matrix $\speciv_\jeig$ in the direction $\kv$ is locally converted to all other modes $\keig$ and directions. Lastly, the Dirac measure in the integral indicates that scattering occurs with possible mode conversion whenever for some directions $\vp$ and $\kv$ one has $\eigl_\keig(\xv,\vp)=\eigl_\jeig(\xv,\kv)$. This includes the situation when no mode conversion occurs, $\jeig=\keig$, and $\eigl_\jeig$ has locally the same value for possibly two different directions. For an isotropic, lossless background medium with $\Kv_\iref(\xg)=\diag(\permitj_\iref(\xv)\Id,\permeaj_\iref(\xv)\Id)$ for example, one has three modes each of multiplicity $2$ with eigenvalues $\eigl_0=0$ (non propagative), $\eigl_1(\xv,\kv)=+\cel_\iref(\xv)\norm{\kv}$, and $\eigl_2(\xv,\kv)=-\cel_\iref(\xv)\norm{\kv}$, where $\cel_\iref=1/\sqrt{\permitj_\iref\permeaj_\iref}$ \cite{RYZ96}. Our results \eqref{eq:Liouville} and \eqref{eq:RTE} generalize \cite{RYZ96} to general bianisotropic dielectric media, and \cite{FAN07b,FAN07c} to the situation where these media are in addition heterogeneous and dissipative.

We now turn to the detailed analysis of the high-frequency limit $\escale\to 0$ of \eref{eq:RandomMaxwell} and how these results are derived. For that purpose, we use a classical pseudo-differential calculus, which also contributes to significantly simplify these derivations.

\section{Transport of high-frequency electromagnetic waves in bianisotropic media}\label{sec:Transport}

In this section the transport equations \eqref{eq:Liouville} for bianisotropic dielectric media are first derived. They describe the evolution of the high-frequency electromagnetic energy density without random fluctuations of the optical response. As illustrated in \sref{sec:WKB}, the high-frequency limit $\escale\rightarrow 0$ in the previous setting shall be derived for quadratic observables of the electromagnetic field $\uv$. More particularly, we introduce the spatio-temporal Wigner transform of that field and its high-frequency limit, \emph{i.e.} its Wigner measure, as in \emph{e.g.} \cite{AKI21}. This is because the spatial and temporal scales in Maxwell's system \eqref{eq:RandomMaxwell} play symmetric roles, and their oscillations should be accounted for altogether. The main objective here is to outline the (formal) mathematical tools used for the derivation of the Wigner measure's properties. They will prove useful in the subsequent~\sref{sec:RTE} where the effects of the random fluctuations of the optical response will be considered. The analysis is derived from~\cite{GER97}, where first-order hyperbolic systems with constant and slowly varying coefficients are addressed, \cite{AKI12}, where arbitrary order hyperbolic systems with slowly varying coefficients are addressed, and \cite{AKI21}, where time-varying random media and consistent pseudo-differential calculus and Wigner transforms with a single quantization are used. In this way we first recall in \sref{sec:WignerPDO} how Wigner transforms are linked to semi-classical pseudo-differential operators \cite{MAR02,ZWO12}. Then writing \eqref{eq:RandomMaxwell} as a semi-classical operator applied to a rescaled version of the electromagnetic field $\uv$ in \sref{sec:PDO-wave-eq}, the dispersion properties of the Wigner measure (related to the Stokes parameters of the electromagnetic waves) are derived in~\sref{sec:dispersion}, and the transport equations \eqref{eq:Liouville} are obtained~\sref{sec:transport}.

\subsection{Semi-classical operators and Wigner measure}\label{sec:WignerPDO}

From now on let us introduce the space-time variable $\xt=(\xg,t)\in\domain\times\Rset$ and its dual variable $\kw=(\kg,\wg)\in\Rset^4$ in the wave vector-frequency Fourier domain. Let $\obsv$ be a smooth, compactly supported real matrix-valued function of both the space-time variable $\xt$ and impulse variable $\kw$. For a vector field $\uv$ in $\smash{L^2(\Rset^4)}$ endowed with the scalar product $\smash{\scald{\uv,\vv}}=\smash{\int_{\Rset^4}\uv(\xt)\cdot\cjg{\vv}(\xt)\,\dd\xt}$ where $\smash{\cjg{\vv}}$ stands for complex conjugation, consider the (semiclassical) pseudo-differential operator:
\begin{equation}
\obsv^\vartheta(\xt,\escale\Dx)\uv(\xt)=\frac{1}{(2\pi)^4}\int_{\Rset^4\times\Rset^4}\iexp^{\ci\kw\cdot(\xt-\yt)}\obsv((1-\vartheta)\xt+\vartheta\yt,\escale\kw)\uv(\yt)\,\dd\yt\dd\kw\,,
\end{equation}
for $\vartheta\in[0,1]$. This parameter defines the so-called quantization of the operator $\obsv^\vartheta$. The case $\vartheta=0$ corresponds to the standard quantization. It is simply denoted by $\smash{\obsv(\xt,\escale\Dx)}$ such that:
\begin{equation}\label{eq:PDO}
\obsv(\xt,\escale\Dx)\uv(\xt)=\frac{1}{(2\pi)^4}\int_{\Rset^4}\iexp^{\ci\kw\cdot\xt}\obsv(\xt,\escale\kw)\TF{\uv}(\kw)\,\dd\kw\,,
\end{equation}
where:
\begin{equation}\label{eq:FT}
\TF{\uv}(\kw)=\int_{\Rset^4}\iexp^{-\ci\kw\cdot\xt}\uv(\xt)\dd\xt
\end{equation}
stands for the space-time Fourier transform of $\uv(\xt)$. The case $\smash{\vartheta=1/2}$ corresponds to Weyl quantization, which is usually denoted by $\smash{\obsv^W(\xt,\escale\Dx)}$. 

Now let $\uv,\vv$ be \alert{tempered} distributions defined on $\smash{\Rset^4}$ and let $\Wignerve^\vartheta[\uv,\vv](\xt,\kw)$ be their Wigner transform \eqref{eq:TWigner} (with $m=4$ here). Then one has the trace formula \cite{GER97}:
\begin{equation}\label{eq:trace}
\scald{\obsv^\vartheta(\xt,\escale\Dx)\uv,\vv}=\trace\int_{\Rset^4\times\Rset^4}\obsv(\xt,\kw)\Wignerve^\vartheta[\uv,\vv](\dd\xt,\dd\kw)\,.
\end{equation}
For a sequence $\smash{\sequence{\uve}}$ uniformly bounded in $\smash{L^2(\Rset^4)}$, one can establish in particular that there exists a positive measure $\Wignerv[\uve]$ such that, up to extracting a subsequence if need be (see \emph{e.g.} \cite[Theorem 5.2]{ZWO12}):
\begin{equation}\label{eq:mes-semi-classique}
\lim_{\escale\rightarrow 0}\scald{\obsv^\vartheta(\xt,\escale\Dx)\uve,\uve}=\trace\int_{\Rset^4\times\Rset^4}\obsv(\xt,\kw)\Wignerv[\uve](\dd\xt,\dd\kw)\,,\quad\forall\obsv\,,
\end{equation}
independently of the quantization $\vartheta$. We thus recover the Wigner measure $\Wignerv[\uve]$ of $\smash{\sequence{\uve}}$ invoked in \sref{sec:WKB}, which can also be interpreted as the weak limit of its self Wigner transform $\smash{\Wignerve^\vartheta[\uve,\uve]:=\Wignerve^\vartheta[\uve]}$. It describes the limit energy density of the sequence $\smash{\sequence{\uve}}$ in the phase space $\smash{\Rset^4_\xt\times\Rset^4_\kw}$. The observable $\obsv(\xt,\kw)$ is used to select any quadratic observable or quantity of interest associated to this energy: the kinetic energy, or the free energy, or the Poynting vector as in \eref{eq:energyHF} for example.

Next, we recall some formulas of pseudo-differential calculus which involve Wigner transforms and semiclassical pseudo-differential operators. In \cite{BAL05} for example, various formal rules of pseudo-differential calculus were given for quantities like $\smash{\Wignerve^{\demi}[\obsv(\xg,\escale\Dxx)\uve,\vve]}$, mixing the quantization chosen for the Wigner transform \eqref{eq:TWigner} ($\vartheta=\demi$) and the standard quantization of the semiclassical operator $\obsv(\xg,\escale\Dxx)$ ($\vartheta=0$). These formulas have been revisited in \cite{AKI21} using the same standard quantization for both the Wigner transforms and pseudo-differential operators, which seems more natural in view of the trace formula \eqref{eq:trace}. They also contribute to significantly simplify the calculations below. They are recalled here, extending slightly the results of \cite{AKI21} for scalar-valued Wigner transforms to matrix-valued Wigner transforms. The proofs are straightforward applications of the proofs in \cite[Appendix B]{AKI21} for the scalar case. For the chosen standard quantization $\vartheta=0$, the Wigner transform \eqref{eq:TWigner} reads:
\begin{equation}\label{eq:TWigner-xt}
\Wignerve[\uve,\vve](\xt,\kw):=\frac{1}{(2\pi)^4}\int_{\Rset^4}\iexp^{\ci\kw\cdot\yt}\uve(\xt-\escale\yt) \adj{\vv}_\escale(\xt)\,\dd\yt\,,
\end{equation}
dropping the superscript $\vartheta$ from now on. The self Wigner transform $\Wignerve[\uve,\uve]$ is denoted by $\Wignerve[\uve]$, as implicitly done in~\eref{eq:mes-semi-classique}. Let $\obsv(\xt,\kw)$ be a smooth matrix-valued function defined on $\smash{\Rset^4_\xt\times\Rset_\kw^4}$ of which derivatives increase only slowly,\footnote{That is, $\obsv$ satisfy for some $m\geq 0$ \cite[Prop. 1.8]{GER97}:
\begin{displaymath}
\norm{\partial_\xt^\alpha\partial_\kw^\beta\obsv(\xt,\kw)}\leq C_{\alpha,\beta}(1+\norm{\kw})^m\,,\quad\forall\alpha,\beta\in\Nset^4\,.
\end{displaymath}}
and recall the notation of \eref{eq:PDO} for the operator $\smash{\obsv(\xt,\escale\Dx)}$ with $\vartheta=0$. Then one has:
\begin{multline}\label{reg1-cal}  
\Wignerve[\obsv(\xt,\escale\Dx )\uve,\vve]= \obsv\left(\xt,\kw\right) \Wignerve[\uve,\vve]-\frac{\escale}{\ci}\bnabla_\xt\obsv\cdot\bnabla_\kw\Wignerve[\uve,\vve] \\
-\frac{\escale}{\ci}\left(\bnabla_\xt\cdot\bnabla_\kw\obsv\right)\Wignerve[\uve,\vve]+\Go(\escale^2)\,, 
\end{multline}
and:
\begin{equation}\label{reg2-cal}  
\begin{split} 
\Wignerve[\uve,\obsv( \xt, \escale \Dx) \vve ] &=\adj{(\obsv(\xt,\kw-\escale \Dx)\Wignerve[\uve,\vve])} \\
&= \Wignerve[\uve,\vve]\adj{\obsv}(\xt,\kw)-\frac{\escale}{\ci}\bnabla_\xt\Wignerve[\uve,\vve]\cdot\bnabla_\kw\adj{\obsv} + \Go(\escale^{2})\,.
\end{split}
\end{equation}
Here $\bnabla_\xt {\boldsymbol A}\cdot\bnabla_\kw {\boldsymbol B}:= \bnabla_{\xg}{\boldsymbol A}\cdot \bnabla_{\kg}{\boldsymbol B}+\partial_t {\boldsymbol A}\partial_\wg {\boldsymbol B}$, and the differential operator $\smash{\escale\Dx}$ within the observable $\obsv$ acts on $\smash{\Wignerve[\uve,\vve]}$ so that:
\begin{displaymath}
\obsv(\xt,\kw -\escale\Dx)\Wignerve[\uve,\vve](\xt,\kw)=\frac{1}{(2\pi)^4}\int_{\Rset^4}\iexp^{\ci\pw\cdot\xt}\obsv(\xt,\kw - \escale\pw)\TF{\Wignerv}_\escale[\uve,\vve](\pw,\kw)\dd\pw\,,
\end{displaymath}
where $\smash{\TF{\Wignerv}_\escale[\uve,\vve](\pw,\kw)}$ is the Fourier transform \eqref{eq:FT} of $\smash{\Wignerv_\escale[\uve,\vve](\xt,\kw)}$ with respect to the space-time variable $\xt$.

\subsection{Maxwell's equations as a semiclassical operator}\label{sec:PDO-wave-eq}

Here \eref{eq:RandomMaxwell} is conveniently written as a pseudo-differential operator for the derivation of the high-frequency regime $\escale\to 0$. We consider slowly varying materials characterized by the Hermitian optical response $\Kv_\iref(\xv)$ which is independent of the small parameter $\escale$ and we ignore its random fluctuations for the time being ($\sigma=0$). Premultiplying it by $\smash{\frac{\escale}{\ci}}$, Maxwell's system \eqref{eq:DampedMaxwell} reads:
\begin{equation}\label{eq:PDOwave}
\vL_\escale(\xt,\escale\Dxt)\uve=\bzero\,,\quad\xt=(\xg,t)\in\domain\times\Rset_+^*\,,
\end{equation}
where $\smash{\Dxt}=\smash{(\Dxx,\Dt)}$ for $\smash{\Dxx=\frac{1}{\ci}\bnabla_\xg}$, $\smash{\iD_t=\frac{1}{\ci}\partial_t}$. The pseudo-differential operator $\vL_\escale$ is given by:
\begin{equation}\label{eq:PDO-L}
\vL_\escale(\xt,\kw)=\wg\Id+\vL_\iref(\xv,\kv)+\frac{\escale}{\ci}\vL_\indL(\xv,\wg)\,, \\
\end{equation}
where the $6\times 6$ matrices $\vL_\iref(\xv,\kv)$ and $\vL_\indL(\xv,\wg)$ are given by \eref{eq:PDOperators}, and the symmetric Maxwell operator $\Maxwell(\kv)$ is given by \eref{eq:MaxwellO}. In \eref{eq:PDO-L} it is assumed that the rescaled susceptibility kernel $\smash{\Kv_d}$ acts as:
\begin{displaymath}
(\Kv_d\star\uve)(\xv,t)=\int_0^t\Kv_d\left(\xv,\frac{\tau}{\escale}\right)\uve(\xv,t-\tau)\dd\tau\,.
\end {displaymath}
A similar model is adopted in \emph{e.g.} \cite{LEW65}. Also $\smash{\uve(\xv,t)}$ stands for the electromagnetic field satisfying \eref{eq:PDOwave} for highly oscillating initial conditions characterized by the small scale $\escale$. We derive its Wigner measure and its evolution properties inside $\domain$ in the subsequent subsections, ignoring boundary effects for now. They can be addressed as in \cite{AKI12} for example, at least for the hyperbolic-elliptic set.

\begin{remark}\label{remark1}
All derivations involving the Wigner transforms and Wigner measures of $\smash{\sequence{\uve}}$ in the subsequent \sref{sec:dispersion} and \sref{sec:transport} are formal. They can be made rigorous as follows. For all $\xt\in \domain \times \smash{\Rset^*_+}$, let ${\mathcal U}$ be an open set such that $\overline{\mathcal U} \subset  \domain \times\smash{\Rset^*_+}$, $\xt \in{\mathcal  U}$, and let $\varphi \in \smash{{\mathcal C}^{\infty}_0(\domain \times \Rset^*_+)}$ with $\varphi \equiv 1$ on $\overline{\mathcal U}$. Multiplying \eref{eq:PDOwave} by $\varphi$, we obtain an alternative equation where $\smash{\uve}$ is replaced by $\varphi\smash{\uve}$, with additional terms. The Wigner measures related to these additional terms are concentrated outside $\overline{\mathcal U}$ because of the semiclassical pseudo-differential calculus \cite[Theorem 4.24]{ZWO12}. Besides, it can be shown that $\Wignerv[\varphi\uve](\xt,\kw)$ is actually independent of $\varphi$ chosen as at the beginning of this remark. In the sequel we shall define the Wigner measure of $\smash{\sequence{\uve}}$, $\smash{\Wignerv[\uve]}(\xt,\kw)$ for $\xt\in \domain \times\smash{\Rset^*_+}$, $\kw\in \Rset^4 \setminus \{\kv =\bzero\}$, as $\smash{\Wignerv[\varphi\uve]}(\xt,\kw)$, with this choice of $\varphi$.
\end{remark}

\subsection{Dispersion properties}\label{sec:dispersion}

The foregoing pseudo-differential calculus and space-time Wigner transform of \sref{sec:WignerPDO} are now used with Maxwell's system (\ref{eq:PDOwave}). Computing the space-time Wigner transform~\eqref{eq:TWigner-xt} of $\uve$ from \eref{eq:PDOwave}, yields:
\begin{equation}\label{eq:step00}
\Wignerve\left[\vL_\escale(\xt,\escale\Dx_{\xt})\uve,\uve\right] - \Wignerve\left[\uve,\vL_\escale(\xt,\escale\Dx_{\xt})\uve\right] = \bzero\,.
\end{equation}
However, invoking the rules \eqref{reg1-cal} and \eqref{reg2-cal} \alert{where it is assumed that $\vL_\iref(\xv,\kv)$ and $\vL_\indL(\xv,\wg)$ have derivatives with respect to $\xv,\kv$, and $\wg$ which increase only slowly, in line with \cite[Prop. 1.8]{GER97}}, we get:
\begin{multline}\label{eq:step0}
\Wignerve[\vL_\escale(\xt,\escale\Dx_{\xt})\uve,\uve] = \left(\wg\Id+\vL_\iref+\frac{\escale}{\ci}\vL_\indL\right)\Wignerve[\uve] - \frac{\escale}{\ci}\bnabla_\xv\vL_\iref\cdot\bnabla_\kv\Wignerve[\uve]  \\
- \frac{\escale}{\ci}(\bnabla_\xv \cdot \bnabla_\kv\vL_\iref)\Wignerve[\uve]+\Go(\escale^2)\,,
\end{multline}
and:
\begin{multline}\label{eq:step0adj}   
\Wignerve[\uve,\vL_\escale(\xt,\escale\Dx_{\xt})\uve ] =\Wignerve[\uve]\left(\wg\Id+\adj{\vL}_0-\frac{\escale}{\ci}\adj{\vL}_\indL\right)-\frac{\escale}{\ci}\partial_t\Wignerve[\uve] \\
-\frac{\escale}{\ci}\bnabla_\xv\Wignerve[\uve]\cdot\bnabla_\kv\adj{\vL}_0 + \Go(\escale^{2})\,.
\end{multline}

Considering the leading-order term, one obtains as $\escale\rightarrow 0$ (see also for instance \cite[Theorem 5.3]{ZWO12}):
\begin{equation}\label{eq:step01}
\vL_0(\xv,\kv)\Wignerv(\xt,\kw)-\Wignerv(\xt,\kw)\adj{\vL}_0(\xv,\kv)=\bzero
\end{equation}
for the Wigner measure $\Wignerv:=\Wignerv[\uve]$ (independent of $\escale$) of the sequence $\smash{\sequence{\uve}}$ (dependent of $\escale$) given by~\eref{eq:mes-semi-classique} in phase space $(\xt,\kw)\in\domain\times\smash{\Rset_+^*}\times\smash{\Rset^4_\kw}\setminus\{\kg=\bzero\}$. Again, the notation of \cite[p. 330]{GER97} is used and the left-hand side in \eref{eq:step01} above is clearly independent of $\escale$. The Wigner measure is a non-negative, Hermitian matrix. Besides, reminding \eref{eq:spectralL0} one deduces from \erefs{eq:step0}{eq:step01}:
\begin{equation}
\Wignerv=\sum_{\jeig}\delta(\wg+\eigl_\jeig)\Wignerv_\jeig\,,
\end{equation}
where $\Wignerv_\jeig=\proj_\jeig\Wignerv=\Wignerv\adj{\proj}_\jeig=\proj_\jeig\Wignerv\adj{\proj}_\jeig$; this is \eref{eq:WignerProj}. We note that here we do not consider eigenvalues crossing, when their multiplicities possibly change. We leave this case for future works. It is considered for example in \cite{FER02} for the Schr\"odinger equation.

\begin{remark}
As noticed in \sref{sec:MainResults}, the non propagative mode $\smash{\eigl_0}=0$ (for $\jeig=\text{"0"}$) is always an eigenvalue of $\smash{\vL_\iref}$ with multiplicity $2$, with the eigenvectors $\smash{\eiga_1}=\smash{(\hkg,\bzero)}$ and $\smash{\eiga_2}=\smash{(\bzero,\hkg)}$ for $\kv\neq\bzero$.
Recall that the optical response $\Kv_\iref$ is positive definite and let $\boldsymbol{\mathcal K}$ be the $2\times 2$ matrix with elements $\smash{{\mathcal K}_{jk}}=\smash{\scal{\eiga_j,\eiga_k}}$, $j,k=1,2$. One has $\boldsymbol{\mathcal K}=\adj{\boldsymbol Q}{\boldsymbol Q}$ where:
\begin{displaymath}
{\boldsymbol Q}=\begin{bmatrix} \sqrt{\hat{\permitj}_\iref} & \frac{\hat{\magnetoj}_\iref}{ \sqrt{\hat{\permitj}_\iref}} \\ 0 & \sqrt{\hat{\permeaj}_\iref-\frac{\norm{\hat{\magnetoj}_\iref}^2}{\hat{\permitj}_\iref}} \end{bmatrix}\,,
\end{displaymath}
with $\smash{\hat{\permitj}_\iref(\xv,\kv)}=\smash{\adj{\hkv}\permitv_\iref(\xv)\hkv}$, $\smash{\hat{\permeaj}_\iref(\xv,\kv)}=\smash{\adj{\hkv}\permeav_\iref(\xv)\hkv}$, and $\smash{\hat{\magnetoj}_\iref(\xv,\kv)}=\smash{\adj{\hkv}\magnetov_\iref(\xv)\hkv}$ for $\kv\neq\bzero$. Then the orthonormal eigenvectors with respect to the scalar product $\scal{\cdot,\cdot}$ are $\smash{\eigb_0^j}={\boldsymbol Q}^{-1}\eiga_j$ since $\smash{\adj{\eiga}_j\eiga_k}=\smash{\delta_{jk}}$, yielding:
\begin{equation}\label{eq:eigenv0}
\eigb_0^{(1)}(\hkv)=\frac{1}{\sqrt{\hat{\permitj}_\iref}}\begin{pmatrix} \hkv \\ \bzero \end{pmatrix}\,,\quad\eigb_0^{(2)}(\hkv)=\frac{1}{\sqrt{\hat{\permitj}_\iref\hat{\permeaj}_\iref-\norm{\hat{\magnetoj}_\iref}^2}}\begin{pmatrix} -\frac{\hat{\magnetoj}_\iref}{\sqrt{\hat{\permitj}_\iref}}\hkv \\ \sqrt{\hat{\permitj}_\iref}\,\hkv \end{pmatrix}\,.
\end{equation}
\end{remark}

\subsection{Evolution properties}\label{sec:transport}

Gathering the first order terms proportional to $\escale$ in \eref{eq:step00} one obtains in the limit $\escale\to 0$:
\begin{multline}\label{eq:step1}
\partial_t\Wignerv + \bnabla_\xv\Wignerv\cdot\bnabla_\kv\adj{\vL}_0 - \bnabla_\xv\vL_0\cdot\bnabla_\kv\Wignerv - (\bnabla_\xv \cdot \bnabla_\kv\vL_0)\Wignerv +\vL_\indL\Wignerv+\Wignerv\adj{\vL}_\indL=\bzero\,.
\end{multline}
Next we multiply \eref{eq:step1} on the left by $\proj_\jeig$ and on the right by $\adj{\proj}_\jeig$. From the property of projectors $\proj_\jeig\proj_\keig=\delta_{\jeig\keig}\proj_\jeig$ we first derive:
\begin{displaymath}
\begin{split}
\proj_\jeig\partial_{\xj_j}\vL_\iref &=\proj_\jeig\sum_\keig\left((\partial_{\xj_j}\eigl_\keig)\proj_\keig+\eigl_\keig(\partial_{\xj_j}\proj_\keig)\right) \\
&=(\partial_{\xj_j}\eigl_\jeig)\proj_\jeig+(\partial_{\xj_j}\proj_\jeig)(\eigl_\jeig-\vL_\iref)\,.
\end{split}
\end{displaymath}
The same holds for partial derivatives with respect to $\kj_j$. Besides:
\begin{displaymath}
\begin{split}
(\partial_{\kj_j}\Wignerv)\adj{\proj}_\jeig &=\partial_{\kj_j}(\Wignerv\adj{\proj}_\jeig)-\Wignerv(\partial_{\kj_j}\adj{\proj}_\jeig) \\
&=\partial_{\kj_j}\Wignerv_\jeig-\Wignerv\adj{(\partial_{\kj_j}\proj_\jeig)}\,
\end{split}
\end{displaymath}
where again the same holds for partial derivatives with respect to $\xj_j$. Therefore (with summation over the repeated index $j$):
\begin{displaymath}
\begin{split}
\proj_\jeig(\partial_{\xj_j}\vL_\iref)(\partial_{\kj_j}\Wignerv)\adj{\proj}_\jeig &=(\partial_{\xj_j}\eigl_\jeig)(\partial_{\kj_j}(\proj_\jeig\Wignerv_\jeig)-(\partial_{\kj_j}\proj_\jeig)\Wignerv_\jeig-\Wignerv_\jeig\adj{(\partial_{\kj_j}\proj_\jeig)}) \\
&\quad+(\partial_{\xj_j}\proj_\jeig)(\eigl_\jeig-\vL_\iref)(\partial_{\kj_j}\Wignerv_\jeig-\Wignerv\adj{(\partial_{\kj_j}\proj_\jeig)}) \\
&=(\partial_{\xj_j}\eigl_\jeig)(\partial_{\kj_j}\Wignerv_\jeig-(\partial_{\kj_j}\proj_\jeig)\Wignerv_\jeig-\Wignerv_\jeig\adj{(\partial_{\kj_j}\proj_\jeig)}) \\
&\quad-(\partial_{\xj_j}\proj_\jeig)((\partial_{\kj_j}(\eigl_\jeig-\vL_\iref))\Wignerv_\jeig+(\eigl_\jeig-\vL_\iref)\Wignerv\adj{(\partial_{\kj_j}\proj_\jeig)})
\end{split}
\end{displaymath}
since $(\eigl_\jeig-\vL_\iref)\Wignerv_\jeig=\bzero$. Because $\partial_{\kj_j}$ and $\partial_{\xj_j}$ play symmetric roles, one also obtains:
\begin{displaymath}
\begin{split}
\proj_\jeig(\partial_{\xj_j}\Wignerv)(\partial_{\kj_j}\adj{\vL}_\iref)\adj{\proj}_\jeig &= \adj{(\proj_\jeig(\partial_{\kj_j}\vL_\iref)(\partial_{\xj_j}\Wignerv)\adj{\proj}_\jeig)}\\
&=(\partial_{\kj_j}\eigl_\jeig)(\partial_{\xj_j}\Wignerv_\jeig-(\partial_{\xj_j}\proj_\jeig)\Wignerv_\jeig-\Wignerv_\jeig\adj{(\partial_{\xj_j}\proj_\jeig)}) \\
&\quad-(\Wignerv_\jeig\partial_{\xj_j}(\eigl_\jeig-\adj{\vL}_\iref)+(\partial_{\xj_j}\proj_\jeig)\Wignerv(\eigl_\jeig-\adj{\vL}_\iref))\adj{(\partial_{\kj_j}\proj_\jeig)}\,.
\end{split}
\end{displaymath}
Gathering these expressions, \eref{eq:step1} yields:
\begin{multline}\label{eq:step2}
\partial_t\Wignerv_\jeig +\{\eigl_\jeig,\Wignerv_\jeig\} + (\bnabla_\xv\eigl_\jeig\cdot\bnabla_\kv\proj_\jeig - \bnabla_\xv\proj_\jeig\cdot\bnabla_\kv\vL_\iref-\proj_\jeig\bnabla_\xv \cdot \bnabla_\kv\vL_0)\Wignerv_\jeig \\
 - \Wignerv_\jeig\adj{(\bnabla_\kv\eigl_\jeig\cdot\bnabla_\xv\proj_\jeig - \bnabla_\kv\proj_\jeig\cdot\bnabla_\xv\vL_\iref)} + \vL_\jeig\Wignerv_\jeig+\Wignerv_\jeig\adj{\vL}_\jeig=\bzero\,,
\end{multline}
where $\{\eigl_\jeig,\Wignerv_\jeig\}=\bnabla_\kv\eigl_\jeig\cdot\bnabla_\xv\Wignerv_\jeig-\bnabla_\xv\eigl_\jeig\cdot\bnabla_\kv\Wignerv_\jeig$ and $\smash{\vL_\jeig}=\smash{\proj_\jeig\vL_\indL}$.

Next consider the matrix $\Mv_\jeig=\partial_{\xj_j}\eigl_\jeig\partial_{\kj_j}\proj_\jeig - \partial_{\xj_j}\proj_\jeig\partial_{\kj_j}\vL_\iref-\proj_\jeig(\partial_{\xj_j}\partial_{\kj_j}\vL_0)$; then:
\begin{displaymath}
\begin{split}
\Mv_\jeig &=\partial_{\kj_j}[(\partial_{\xj_j}\eigl_\jeig)\proj_\jeig]-(\partial_{\xj_j}\partial_{\kj_j}\eigl_\jeig)\proj_\jeig-\partial_{\xj_j}(\proj_\jeig\partial_{\kj_j}\vL_\iref) \\
&=\partial_{\kj_j}[\cancel{\partial_{\xj_j}(\eigl_\jeig\proj_\jeig)}-\eigl_\jeig\partial_{\xj_j}\proj_\jeig]-(\partial_{\xj_j}\partial_{\kj_j}\eigl_\jeig)\proj_\jeig-\partial_{\xj_j}[\cancel{\partial_{\kj_j}(\proj_\jeig\vL_\iref)}-(\partial_{\kj_j}\proj_\jeig)\vL_\iref] \\
&=-(\partial_{\kj_j}\eigl_\jeig)(\partial_{\xj_j}\proj_\jeig) - (\partial_{\kj_j}\partial_{\xj_j}\proj_\jeig)(\eigl_\jeig-\vL_\iref)-(\partial_{\xj_j}\partial_{\kj_j}\eigl_\jeig)\proj_\jeig+(\partial_{\kj_j}\proj_\jeig)(\partial_{\xj_j}\vL_\iref)\,.
\end{split}
\end{displaymath}
Letting (because again $(\eigl_\jeig-\vL_\iref)\Wignerv_\jeig=\bzero$):
\begin{equation}
\Nv_\jeig=\bnabla_\kg\proj_\jeig\cdot\bnabla_\xg\vL_\iref-\bnabla_\kg\eigl_\jeig\cdot\bnabla_\xg\proj_\jeig-\demi(\bnabla_\xg\cdot\bnabla_\kg\eigl_\jeig)\Id\,,
\end{equation}
\eref{eq:step2} finally reads:
\begin{equation}\label{eq:step3}
\partial_t\Wignerv_\jeig +\{\eigl_\jeig,\Wignerv_\jeig\} + (\vL_\jeig+\Nv_\jeig)\Wignerv_\jeig + \Wignerv_\jeig\adj{(\vL_\jeig+\Nv_\jeig)} =\bzero\,.
\end{equation}

Alternatively, one may wish to expand the Wigner measure $\Wignerv$ of \eref{eq:WignerProj} onto the eigenvectors $\eigb_\jeig$ as:
\begin{displaymath}
\Wignerv=\sum_{\jeig}\delta(\wg+\eigl_\jeig)\Wignerv_\jeig=\sum_{\jeig}\delta(\wg+\eigl_\jeig)\eigb_\jeig\speciv_\jeig\adj{\eigb}_\jeig\,,
\end{displaymath}
where $\speciv_\jeig=\adj{\eigc}_\jeig\Wignerv_\jeig\eigc_\jeig$. The evolution properties of the so-called specific intensity, or coherence matrices $\speciv_\jeig$ are derived multiplying \eref{eq:step3} by $\adj{\eigc}_\jeig$ on the left-hand side, and by $\eigc_\jeig$ on the right-hand side. Firstly:
\begin{displaymath}
\begin{split}
\adj{\eigc}_\jeig(\partial_{\kj_j}\eigl_\jeig)(\partial_{\xj_j}\Wignerv_\jeig)\eigc_\jeig &=(\partial_{\kj_j}\eigl_\jeig)[\partial_{\xj_j}\speciv_\jeig-(\partial_{\xj_j}\adj{\eigc}_\jeig)\Wignerv_\jeig\eigc_\jeig-\adj{\eigc}_\jeig\Wignerv_\jeig\partial_{\xj_j}\eigc_\jeig] \\
&=(\partial_{\kj_j}\eigl_\jeig)[\partial_{\xj_j}\speciv_\jeig-(\partial_{\xj_j}\adj{\eigc}_\jeig)\eigb_\jeig\speciv_\jeig-\speciv_\jeig\adj{\eigb}_\jeig\partial_{\xj_j}\eigc_\jeig] \\
&=(\partial_{\kj_j}\eigl_\jeig)[\partial_{\xj_j}\speciv_\jeig + \adj{\eigc}_\jeig(\partial_{\xj_j}\eigb_\jeig)\speciv_\jeig + \speciv_\jeig(\partial_{\xj_j}\adj{\eigb}_\jeig)\eigc_\jeig]
\end{split}
\end{displaymath}
owing to the orthonormality condition $\adj{\eigc}_\jeig\eigb_\keig=\delta_{\jeig\keig}\Id_A$ which yields $(\partial_{\xj_j}\adj{\eigc}_\jeig)\eigb_\keig=-\adj{\eigc}_\jeig(\partial_{\xj_j}\eigb_\keig)$ (and a similar property with $\partial_{\kj_j}$). Likewise:
\begin{displaymath}
\adj{\eigc}_\jeig(\partial_{\xj_j}\eigl_\jeig)(\partial_{\kj_j}\Wignerv_\jeig)\eigc_\jeig = (\partial_{\xj_j}\eigl_\jeig)[\partial_{\kj_j}\speciv_\jeig + \adj{\eigc}_\jeig(\partial_{\kj_j}\eigb_\jeig)\speciv_\jeig + \speciv_\jeig(\partial_{\kj_j}\adj{\eigb}_\jeig)\eigc_\jeig]\,,
\end{displaymath}
and consequently:
\begin{displaymath}
\adj{\eigc}_\jeig\{\eigl_\jeig,\Wignerv_\jeig\}\eigc_\jeig=\{\eigl_\jeig,\speciv_\jeig\} + \adj{\eigc}_\jeig\{\eigl_\jeig,\eigb_\jeig\}\speciv_\jeig + \speciv_\jeig\{\eigl_\jeig,\adj{\eigb}_\jeig\}\eigc_\jeig\,.
\end{displaymath}
Secondly:
\begin{displaymath}
\begin{split}
\adj{\eigc}_\jeig(\partial_{\kj_j}\proj_\jeig)(\partial_{\xj_j}\vL_\iref)\Wignerv_\jeig\eigc_\jeig &= [\partial_{\kj_j}(\adj{\eigc}_\jeig\proj_\jeig)-(\partial_{\kj_j}\adj{\eigc}_\jeig)\proj_\jeig](\partial_{\xj_j}\vL_\iref)\eigb_\jeig\speciv_\jeig \\
&=(\partial_{\kj_j}\adj{\eigc}_\jeig)(\Id-\proj_\jeig)[\partial_{\xj_j}(\vL_\iref\eigb_\jeig)-\vL_\iref\partial_{\xj_j}\eigb_\jeig]\speciv_\jeig \\
&=(\partial_{\kj_j}\adj{\eigc}_\jeig)(\Id-\proj_\jeig)[\partial_{\xj_j}(\eigl_\jeig\eigb_\jeig)-\vL_\iref\partial_{\xj_j}\eigb_\jeig]\speciv_\jeig \\
&=(\partial_{\kj_j}\adj{\eigc}_\jeig)(\Id-\proj_\jeig)(\eigl_\jeig-\vL_\iref)(\partial_{\xj_j}\eigb_\jeig)\speciv_\jeig \\
&=(\partial_{\kj_j}\adj{\eigc}_\jeig)(\eigl_\jeig-\vL_\iref)(\partial_{\xj_j}\eigb_\jeig)\speciv_\jeig \\
&=-\adj{\eigc}_\jeig[\partial_{\kj_j}(\eigl_\jeig-\vL_\iref)](\partial_{\xj_j}\eigb_\jeig)\speciv_\jeig\,;
\end{split}
\end{displaymath}
and:
\begin{displaymath}
\begin{split}
\adj{\eigc}_\jeig(\partial_{\kj_j}\eigl_\jeig)(\partial_{\xj_j}\proj_\jeig)\Wignerv_\jeig\eigc_\jeig &= (\partial_{\kj_j}\eigl_\jeig)[\partial_{\xj_j}(\adj{\eigc}_\jeig\proj_\jeig)-(\partial_{\xj_j}\adj{\eigc}_\jeig)\proj_\jeig]\eigb_\jeig\speciv_\jeig \\
&=(\partial_{\kj_j}\eigl_\jeig)(\partial_{\xj_j}\adj{\eigc}_\jeig)(\Id-\proj_\jeig)\eigb_\jeig\speciv_\jeig \\
&=\bzero \,.
\end{split}
\end{displaymath}
Thirdly:
\begin{displaymath}
\begin{split}
\adj{\eigc}_\jeig\{\eigl_\jeig,\eigb_\jeig\}-\adj{\eigc}_\jeig[\partial_{\kj_j}(\eigl_\jeig-\vL_\iref)]\partial_{\xj_j}\eigb_\jeig &=\adj{\eigc}_\jeig[(\partial_{\kj_j}\vL_\iref)(\partial_{\xj_j}\eigb_\jeig)-(\partial_{\xj_j}\eigl_\jeig)(\partial_{\kj_j}\eigb_\jeig)]\,.
\end{split}
\end{displaymath}
Letting:
\begin{equation}\label{eq:skew-symmatrix}
\begin{split}
\nv_\jeig &= \adj{\eigc}_\jeig(\{\eigl_\jeig,\eigb_\jeig\}+\Nv_\jeig\eigb_\jeig) \\
&=\adj{\eigc}_\jeig(\bnabla_\kv\vL_\iref\cdot\bnabla_\xv\eigb_\jeig-\bnabla_\xv\eigl_\jeig\cdot\bnabla_\kv\eigb_\jeig) - \demi(\bnabla_\xg\cdot\bnabla_\kg\eigl_\jeig)\Id_A\,,
\end{split}
\end{equation}
one can show that $\nv_\jeig$ is skew-symmetric owing to \eref{eq:PDOperators}. Indeed:
\begin{displaymath}
\begin{split}
\adj{\eigc}_\jeig(\partial_{\kj_j}\vL_\iref)\eigb_\jeig &= \adj{\eigc}_\jeig\Kv_\iref^{-1}(\xv)(\partial_{\kj_j}\Maxwell(\kv))\eigb_\jeig \\
&= \adj{\eigb}_\jeig(\partial_{\kj_j}\Maxwell)\eigb_\jeig \\
&= (\partial_{\kj_j}\eigl_\jeig)\Id_A\,.
\end{split}
\end{displaymath}
Then $(\partial_{\xj_j}\adj{\eigb}_\jeig)(\partial_{\kj_j}\Maxwell)\eigb_\jeig+\adj{\eigb}_\jeig(\partial_{\kj_j}\Maxwell)(\partial_{\xj_j}\eigb_\jeig)=(\partial_{\xj_j}\partial_{\kj_j}\eigl_\jeig)\Id_A$, or:
\begin{equation}\label{eq:step31}
\begin{aligned}[b]
\adj{\eigc}_\jeig(\partial_{\kj_j}\vL_\iref)(\partial_{\xj_j}\eigb_\jeig) &=(\partial_{\xj_j}\partial_{\kj_j}\eigl_\jeig)\Id_A - \adj{(\adj{\eigb}_\jeig(\partial_{\kj_j}\Maxwell)(\partial_{\xj_j}\eigb_\jeig))} \\
&=(\partial_{\xj_j}\partial_{\kj_j}\eigl_\jeig)\Id_A - \adj{(\adj{\eigc}_\jeig(\partial_{\kj_j}\vL_\iref)(\partial_{\xj_j}\eigb_\jeig))}\,;
\end{aligned}
\end{equation}
besides:
\begin{equation}\label{eq:step32}
\begin{aligned}[b]
(\partial_{\xj_j}\eigl_\jeig)\adj{\eigc}_\jeig\partial_{\kj_j}\eigb_\jeig &=-(\partial_{\xj_j}\eigl_\jeig)(\partial_{\kj_j}\adj{\eigc}_\jeig)\eigb_\jeig \\
&=-(\partial_{\xj_j}\eigl_\jeig)(\partial_{\kj_j}\adj{\eigb}_\jeig)\eigc_\jeig\,;
\end{aligned}
\end{equation}
thus one concludes with (\ref{eq:step31}) and (\ref{eq:step32}) that $\nv_\jeig=-\adj{\nv}_\jeig$. Therefore, setting $\smash{\vl_\jeig}=\smash{\adj{\eigc}_\jeig\vL_\indL\eigb_\jeig}$, \eref{eq:step3} for the specific intensities $\speciv_\jeig$ finally reads:
\begin{equation}\label{eq:step4}
\partial_t\speciv_\jeig+\{\eigl_\jeig,\speciv_\jeig\}  + \vl_\jeig\speciv_\jeig + \speciv_\jeig\adj{\vl}_\jeig + \nv_\jeig\speciv_\jeig - \speciv_\jeig\nv_\jeig = \bzero\,.
\end{equation}
One can check in particular that for the eigenspace (\ref{eq:eigenv0}) associated with the null eigenvalue ($\jeig=\text{"0"}$), one has $\nv_0=\adj{\eigc}_0\bnabla_\kv\vL_\iref\cdot\bnabla_\xv\eigb_0=\adj{\eigb}_0\bnabla_\kv\Maxwell(\kv)\cdot\bnabla_\xv\eigb_0=\bzero$.

The geometrical structure of transport equations (\ref{eq:step4}) is analyzed in \emph{e.g.} \cite{LES15}. Here we show how solving \eref{eq:step4} amounts to track $\smash{\speciv_\jeig}$ on its bicharacteristic rays in $\domain\times\Rset^3$. Consider the bicharacteristic (Hamiltonian) equations associated to $\eigl_\jeig(\xv,\kv)$:
\begin{equation}\label{eq:bicharacteristics}
\begin{split}
\frac{\dd\xv_\jeig}{\dd t} &=\bnabla_\kv\eigl_\jeig(\xv_\jeig(t),\kv_\jeig(t))\,,\\
\frac{\dd\kv_\jeig}{\dd t} &=-\bnabla_\xv\eigl_\jeig(\xv_\jeig(t),\kv_\jeig(t))\,,
\end{split}
\end{equation}
with initial conditions $\smash{\xv_\jeig(0)}=\smash{\xv_0}$ and $\smash{\kv_\jeig(0)}=\smash{\kv_0}$ lying in the support of the Wigner measure $\smash{\Wignerv_I}$ of the initial conditions $\smash{\uve(\cdot,t=0)}$. Also introduce $\smash{\Omv_\jeig}=\vl_\jeig+\nv_\jeig$ and the $A\times A$ matrix $\smash{\Rv_\jeig}$ which is such that:
\begin{equation}\label{eq:rotation}
\frac{\dd\Rv_\jeig}{\dd t} =-\Omv_\jeig(\xv_\jeig(t),\kv_\jeig(t))\Rv_\jeig(t)\,,\quad\Rv_\jeig(0)=\Id_A\,.
\end{equation}
Then: 
\begin{equation*}
\begin{split}
\Omv_\jeig\speciv_\jeig+\speciv_\jeig\adj{\Omv}_\jeig &=-\frac{\dd\Rv_\jeig}{\dd t}\Rv^{-1}_\jeig\speciv_\jeig - \speciv_\jeig(\adj{\Rv}_\jeig)^{-1}\frac{\dd\adj{\Rv}_\jeig}{\dd t} \\
&=-\frac{\dd\Rv_\jeig}{\dd t}\tilde{\speciv}_\jeig\adj{\Rv}_\jeig - \Rv_\jeig\tilde{\speciv}_\jeig\frac{\dd\adj{\Rv}_\jeig}{\dd t}
\end{split}
\end{equation*}
if one lets $\tilde{\speciv}_\jeig=\Rv^{-1}_\jeig\speciv_\jeig(\adj{\Rv}_\jeig)^{-1}$, and owing to \eref{eq:bicharacteristics} the transport equation \eqref{eq:Liouville} reads:
\begin{equation*}
\frac{\dd\speciv_\jeig}{\dd t} - \frac{\dd\Rv_\jeig}{\dd t}\tilde{\speciv}_\jeig\adj{\Rv}_\jeig - \Rv_\jeig\tilde{\speciv}_\jeig\frac{\dd\adj{\Rv}_\jeig}{\dd t}=\bzero\,,
\end{equation*}
or:
\begin{equation}
\frac{\dd}{\dd t}\tilde{\speciv}_\jeig(\xv_\jeig(t),\kv_\jeig(t),t)=\bzero\,.
\end{equation}
Consequently:
\begin{equation}
\speciv_\jeig(\xv_\jeig(t),\kv_\jeig(t),t)=\Rv_\jeig(t)\speciv^I_\jeig(\xv_0,\kv_0)\adj{\Rv}_\jeig(t)
\end{equation}
where $\smash{\speciv^I_\jeig}=\smash{\adj{\eigc}_\jeig\Wignerv_I\eigc_\jeig}$, and one comes down to solving \erefs{eq:bicharacteristics}{eq:rotation} in order to track $\smash{\speciv_\jeig}$ on its bicharacteristics $t\mapsto\smash{(\xv_\jeig(t),\kv_\jeig(t))}$.

\begin{example}\label{example2}
For the Lorentz model with damping (\ref{eq:Lorentz}) one has three eigenvalues $\eigl_0=0$ ($\jeig=\text{"0"}$), $\eigl_+(\kv)=+\cel_\iref\norm{\kv}$ ($\jeig=\text{"+"}$), and $\eigl_-(\kv)=-\cel_\iref\norm{\kv}$ ($\jeig=\text{"$-$"}$), each of multiplicity two ($A=2$), $\smash{\cel_\iref}=\smash{1/\sqrt{\permitj_\iref\permeaj_\iref}}$, with associated eigenvectors \cite{RYZ96}:
\begin{displaymath}
\eigb_0^{(1)}(\hkv)=\frac{1}{\sqrt{\permitj_\iref}}\begin{pmatrix}\hkv \\ \bzero\end{pmatrix}\,,\quad\eigb_0^{(2)}(\hkv)=\frac{1}{\sqrt{\permeaj_\iref}}\begin{pmatrix}\bzero \\ \hkv\end{pmatrix}\,,
\end{displaymath}
and for $a=1,2$:
\begin{equation}\label{eq:eigv-Lorentz}
\eigb_\pm^a(\hkv)=\begin{pmatrix} \displaystyle\frac{\ev_a(\hkv)}{\sqrt{2\permitj_\iref}}\\ \pm\displaystyle\frac{\hkv\times\ev_a(\hkv)}{\sqrt{2\permeaj_\iref}} \end{pmatrix}\,,\quad\ev_1(\hkv)\perp\ev_2(\hkv)\in\hkv^\perp\,,
\end{equation}
such that $\smash{(\hkv,\ev_1(\hkv),\ev_2(\hkv))}$ form a right-handed orthonormal system. Besides $\smash{\vl_0}\smash{\speciv_0}=\bzero$ since $\wg=-\smash{\eigl_0}=0$ on the support of $\smash{\speciv_0}$, and:
\begin{displaymath}
\vl_\pm=\frac{\ci\wg\TF{\permitj}_\indL(\wg)}{2}\Id_2
\end{displaymath}
with $\TF{\permitj}_\indL(\wg)=\smash{\frac{\TF{\permitj}_d(\wg)}{\permitj_\iref}}$, such that for the propagating modes $\smash{\eigl_\pm}(\kv)=\pm\smash{\cel_\iref}\norm{\kg}$:
\begin{equation}
\begin{split}
 \vl_\pm\speciv_\pm+\speciv_\pm\adj{\vl}_\pm &=\Re\mathfrak{e}\{\ci\wg\TF{\permitj}_\indL(\wg)\}\speciv_\pm \\
 &=\frac{\wg_p^2\cel_\iref^2\norm{\kv}^2\Gamma}{(\wg_0^2-\cel_\iref^2\norm{\kv}^2)^2+\cel_\iref^2\norm{\kv}^2\Gamma^2}\speciv_\pm \\
 &=\tilde{\Gamma}(\cel_\iref\norm{\kv})\speciv_\pm\,.
 \end{split}
\end{equation}
Also $\smash{\nv_\pm}=\bzero$ for an homogeneous material. Therefore the solution of \eref{eq:step4} reads:
\begin{equation}
\speciv_\pm(\xv,\kg,t)=\iexp^{-\tilde{\Gamma}(\cel_\iref\abs{\kg})t}\speciv_I(\xv\mp\cel_\iref\hkv t,\kv)
\end{equation}
where $\speciv_I(\xv,\kv)$, $(\xv,\kv)\in\domain\times\Rset^3$, are the initial (incident) specific intensities supported in some subset of the physical domain $\domain$. Note that similar results were obtained in \cite{ATA04,PET03} for memoryless viscoelastic media. Recently a semi-linear radiative transfer model has been derived in media with non linear absorption \cite{KRA23}.
\end{example}

\section{Radiative transfer}\label{sec:RTE}

In this section the radiative transfer equations \eqref{eq:RTE} for a bianistropic dielectric medium are now derived. They describe the evolution of the high-frequency electromagnetic energy density accounting for random fluctuations of the optical response. We resort to the same formal pseudo-differential calculus as recalled in \sref{sec:WignerPDO}. However additional rules are needed to account for the dependency of the optical response on both the slow scale of variation of the background medium, and the fast scale of variation of the random fluctuations. In \sref{sec:RandomMaxwell} below we start by rewriting Maxwell's system as a pseudo-differential operator when these random fluctuations are considered. In \sref{sec:pse-dif-cal} we give the additional rules of pseudo-differential calculus for this situation, and we finally obtain in \sref{sec:multiscale} the radiative transfer equations by a multiscale analysis.

\subsection{Maxwell's equations with random perturbations}\label{sec:RandomMaxwell}

In view of the scattering regime of \erefs{eq:lcor}{eq:sigma}, the rescaled optical response $\Kv(\xv)$ in \eref{eq:RandomMaxwell} is written :
\begin{displaymath}
\Kv(\xv) = \Kv_\iref(\xv)\left[\Id+\sqrt{\escale}\Kper\left(\frac{\xv}{\escale}\right)\right]\,.
\end{displaymath}
The power spectral density matrix $\TF{\coro}$ of the homogeneous random field $(\Kper(\yv),\,\yv\in\Rset^3)$ is given by \eref{eq:Bochner} and satisfies:
\begin{equation}\label{eq:DSP}
\esp{\smash{\TF{\Kper}(\vp)\otimes\TF{\Kper}(\vq)}}=(2\pi)^6\TF{\coro}(\vp)\delta(\vp+\vq)
\end{equation}
for the Fourier transform $\TF{\Kper}$ of $\Kper$; see \eref{eq:FT}. Besides, homogeneity (spatial stationarity) also implies that:
\begin{equation}\label{eq:DSPsym}
\TF{\coroij}_{jklm}(\vp)=\TF{\coroij}_{lmjk}(-\vp)\,, \quad 1\leq j,k,l,m\leq 6\,.
\end{equation}
We remind that in order to preserve the Hermiticity of the actual optical response $\Kv$, it is assumed that $\Kv_\iref(\xv)\Kper(\yv)=\adj{\Kper(\yv)}\Kv_\iref(\xv)$ for all $\xv\in\domain$, $\yv=\frac{\xv}{\escale}$. The pseudo-differential operator $\vL_\escale$ of \eref{eq:PDO-L} is then modified to:
\begin{equation}
\mathcal{L}_\escale\left(\xt,\frac{\xv}{\escale},\kw\right)=\vL_\escale(\xt,\kw) + \sqrt{\escale}\vL_\half\left(\frac{\xv}{\escale},\wg\right)\,,
\end{equation}
where again $\xt=(\xv,t)$, $\kw=(\kv,\wg)$, and $\smash{\vL_\half(\yv,\wg)=\wg\Kper(\yv)}$. This additional term arises from the random fluctuations of the optical response. Since it depends on the fast variable $\yv=\frac{\xv}{\escale}$, the rules \eqref{reg1-cal} and \eqref{reg2-cal} have to be adapted.

\subsection{Rules of pseudo-differential calculus with oscillating coefficients} \label{sec:pse-dif-cal}

The rules \eqref{reg1-cal} and \eqref{reg2-cal} hold true if the Wigner transform $\smash{\Wignerve[\uve,\vve]}$ does not depend on $\smash{\xte}$. If however it depends on this fast variable they have to be modified accordingly. Let $\yt\mapsto {\boldsymbol A}(\yt)$ be a smooth matrix-valued function. Then we have: 
\begin{equation}\label{reg4-cal}
\begin{split}
\Wignerve\left[ {\boldsymbol A}\left(\xte\right) \uve,\vve\right] &=  
 \frac{1}{(2\pi)^4} \int_{\Rset^4}  \iexp^{\ci\xte\cdot\pw} \TF{ {\boldsymbol A}} (\pw)  \Wignerve[ \uve,\vve ] \left(\xt,\kw-\pw\right) \dd\pw \,,   \\  
\Wignerve\left[\uve, {\boldsymbol A}\left(\xte\right)\vve \right] &= \Wignerve\left[\uve,\vve \right]\adj{\boldsymbol A\left(\xte\right)} \,,
\end{split}
\end{equation}
where $ \TF{{\boldsymbol A}}(\pw)$ is the Fourier transform \eqref{eq:FT} of ${\boldsymbol A}(\yt)$. Applying these formulas for a matrix-valued observable $\smash{{\boldsymbol A}(\xte)\obsv(\xt,\kw)}$ yields:
\begin{multline}\label{reg5-cal}
\Wignerve\left[{\boldsymbol A}\left(\xte\right)\obsv(\xt,\escale\Dx)\uve,\vve\right] \\
=\int\frac{\iexp^{\ci\xte\cdot\pw}\dd\pw}{(2\pi)^4}\TF{{\boldsymbol A}}(\pw)\obsv(\xt,\kw-\pw)\Wignere[\uve,\vve]\left(\xt,\kw-\pw\right)+\Go(\escale)\,, 
\end{multline}
and:
\begin{equation}\label{reg5adj-cal}
\Wignerve\left[\uve,{\boldsymbol A}\left(\xte\right)\obsv(\xt,\escale\Dx)\vve\right] = \adj{\left({\boldsymbol A}\left(\xte\right)\obsv(\xt,\kw-\escale\Dx) \Wignerve\left[\uve,\vve\right]\right)} \,,
\end{equation}
owing to \eref{reg1-cal} and \eref{reg2-cal}. The proofs of these formulas are given in \cite[Appendix C]{AKI21} for scalar fields. Their extensions to vector fields $\uve$ and $\vve$ are straightforward. Consequently one has:
\begin{equation}
\begin{split}
\Wignerve\left[\vL_\half\left(\frac{\xv}{\escale},\escale\Dt\right) \uve,\uve\right] &=  
 \frac{\wg}{(2\pi)^3} \int_{\Rset^3}  \iexp^{\ci\frac{\xv}{\escale}\cdot\vp} \TF{\Kper} (\vp) \Wignerve[\uve] (\xt,\kv-\vp,\wg) \dd\vp \,,   \\  
\Wignerve\left[\uve, \vL_\half\left(\frac{\xv}{\escale},\escale\Dt\right)\uve \right] &= (\wg-\escale\Dt)\Wignerve[\uve](\xt,\kw)\adj{\Kper\left(\frac{\xv}{\escale}\right)} \,.
\end{split}
\end{equation}
Considering again the rules (\ref{reg1-cal}) and (\ref{reg2-cal}) one obtains that the Wigner transform $\Wignerve[\uve]$ satisfies now:
\begin{multline}\label{eq:WignerMoyal-random}
\vL_\iref(\xv,\kv)\Wignerve[\uve] -\adj{(\vL_\iref(\xv,\kv-\escale\Dx_\xv)\Wignerve[\uve])} \\
+\sqrt{\escale}\left( \frac{\wg}{(2\pi)^3} \int_{\Rset^3}  \iexp^{\ci\frac{\xv}{\escale}\cdot\vp} \TF{\Kper} (\vp) \Wignerve[\uve] (\xt,\kv-\vp,\wg) \dd\vp - \wg\Wignerve[\uve]\adj{\Kper\left(\frac{\xv}{\escale}\right)}\right) \\
+\frac{\escale}{\ci}\left(\partial_t\Wignerve[\uve]-\bnabla_\xv\vL_\iref\cdot\bnabla_\kv\Wignerve[\uve]-(\bnabla_\xv\cdot\bnabla_\kv\vL_\iref)\Wignerve[\uve]+\bnabla_\xv\Wignerve[\uve]\cdot\bnabla_\kv\adj{\vL}_\iref\right) \\
+\frac{\escale}{\ci}\vL_\indL\Wignerve[\uve] + \frac{\escale}{\ci}\Wignerve[\uve]\adj{\vL}_\indL= \Go(\escale^\frac{3}{2})\,.
\end{multline}

\subsection{Multiscale expansion}\label{sec:multiscale}

Because of the two-scale dependency of $\mathcal{L}_\escale$ in terms of the slow variable $\xv$ and the fast variable $\yv=\frac{\xv}{\escale}$, we expand $\Wignerve[\uve]$ as:
\begin{equation}
\Wignerve[\uve]\left(\xt,\kw\right)=\Wignerv_0(\xt,\kw) + \sqrt{\escale}\Wignerv_\half(\xt,\yv,\kw) + \escale\Wignerv_1(\xt,\yv,\kw) + \po(\escale)\,,
\end{equation}
and replace $\escale\Dx_\xv$ by $\escale\Dx_\xv+\Dx_\yv$ in \eref{eq:WignerMoyal-random}. The $0$--th order equation (\ref{eq:step01}) is recovered for $\Wignerv_0$ which is independent of the fast variable $\yv$, so that it is given by \eref{eq:WignerProj} alike:
\begin{equation}\label{eq:WignerProj1}
\Wignerv_0(\xt,\kv,\wg)=\sum_{\jeig}\delta(\wg+\eigl_\jeig(\xv,\kv))\Wignerv_\jeig(\xt,\kv)\,.
\end{equation}

The $\Go(\escale^\demi)$ half-order terms yield:
\begin{multline}
\vL_\iref(\xv,\kv)\Wignerv_\half(\xt,\yv,\kw) - \adj{(\vL_\iref(\xv,\kv-\Dx_\yv)\Wignerv_\half)}(\xt,\yv,\kw) \\
+ \frac{\wg}{(2\pi)^3} \int_{\Rset^3}  \iexp^{\ci\yv\cdot\vp} \TF{\Kper} (\vp) \Wignerv_0 (\xt,\kv-\vp,\wg) \dd\vp - \wg\Wignerv_0(\xt,\kv,\wg)\adj{\Kper(\yv)} =\bzero\,.
\end{multline}
Taking the Fourier transform with respect to $\yv$ of the above equation yields:
\begin{multline}\label{eq:corrector}
\vL_\iref(\xv,\kv)\TF{\Wignerv}_\half(\xt,\vp,\kv,\wg) - \TF{\Wignerv}_\half(\xt,\vp,\kv,\wg)\adj{\vL}_\iref(\xv,\kv-\vp) \\
+\wg\TF{\Kper} (\vp) \Wignerv_0 (\xt,\kv-\vp,\wg)  - \wg\Wignerv_0(\xt,\kv,\wg)\adj{\TF{\Kper}(-\vp)} =\bzero\,.
\end{multline}
The corrector $\TF{\Wignerv}_\half$ is then expanded as (we remove the $\xt$ and $\wg$ dependences from now on for clarity purposes):
\begin{displaymath}
\TF{\Wignerv}_\half(\vp,\kv)=\sum_{\jeig,\keig}\eigb_\jeig(\kv)\TF{\speciv}_{\jeig\keig}(\vp,\kv)\adj{\eigb}_\keig(\kv-\vp)\,,
\end{displaymath}
where from \eref{eq:corrector} one deduces that:
\begin{displaymath}
\TF{\speciv}_{\jeig\keig}(\vp,\kv)=\adj{\eigc}_\jeig(\kv)\frac{\eigl_\keig(\kv-\vp)\TF{\Kper}(\vp)\Wignerv_\keig(\kv-\vp) - \eigl_\jeig(\kv)\Wignerv_\jeig(\kv)\adj{\TF{\Kper}(-\vp)}}{\eigl_\jeig(\kv)-\eigl_\keig(\kv-\vp)-\ci\theta}\eigc_\keig(\kv-\vp)\,.
\end{displaymath}
Here it is implicitly understood from \eref{eq:WignerProj1} that $\wg=-\eigl_\keig(\kv-\vp)$ on the support of $\Wignerv_\keig(\kv-\vp)$ and that $\wg=-\eigl_\jeig(\kv)$ on the support of $\Wignerv_\jeig(\kv)$, while these terms vanish away from these supports. Also the regularization parameter $\theta\in\Rset$ is introduced to evade the case $\eigl_\jeig(\kv)=\eigl_\keig(\kv-\vp)$ for the time being. It will eventually be sent to $0$ at the end of the derivation. Therefore one has:\footnote{This result should be compared with \cite[Eq. (35)]{FAN07b} or \cite[Eq. (35)]{FAN07c} where no coupling between modes is considered.}
\begin{multline}\label{eq:Whalf}
\TF{\Wignerv}_\half(\vp,\kv)= \\
\sum_{\jeig,\keig}\frac{\eigl_\keig(\kv-\vp)\proj_\jeig(\kv)\TF{\Kper}(\vp)\Wignerv_\keig(\kv-\vp) - \eigl_\jeig(\kv)\Wignerv_\jeig(\kv)\adj{\TF{\Kper}(-\vp)}\adj{\proj_\keig(\kv-\vp)}}{\eigl_\jeig(\kv)-\eigl_\keig(\kv-\vp)-\ci\theta}\,.
\end{multline}

At last, the $\Go(\escale)$ terms yield:
\begin{multline*}
\partial_t\Wignerv_0-\bnabla_\xv\vL_\iref\cdot\bnabla_\kv\Wignerv_0-(\bnabla_\xv\cdot\bnabla_\kv\vL_\iref)\Wignerv_0+\bnabla_\xv\Wignerv_0\cdot\bnabla_\kv\adj{\vL}_\iref + \vL_\indL\Wignerv_0 + \Wignerv_0\adj{\vL}_\indL \\
+\frac{\ci\wg}{(2\pi)^3} \int_{\Rset^3}  \iexp^{\ci\yv\cdot\vp} \TF{\Kper} (\vp) \Wignerv_\half(\yv,\kv-\vp) \dd\vp - \ci\wg\Wignerv_\half(\yv,\kv)\adj{\Kper(\yv)} \\
+\ci\vL_\iref(\kv)\Wignerv_1 + \adj{(\ci\vL_\iref(\kv-\Dx_\yv)\Wignerv_1)}= \bzero\,.
\end{multline*}
We plug \eref{eq:Whalf} into the above equation, multiply it on the left by $\smash{\adj{\eigc}_\jeig(\kv)}$ and on the right by $\smash{\eigc_\jeig(\kv)}$, and finally take the average $\esp{\cdot}$ with respect to the random perturbations $\Kper$ with due consideration of \eref{eq:DSP}. The terms in $\Wignerv_1$ vanish once they are averaged since $\smash{\esp{\Wignerv_1}}=\bzero$ for $\smash{\esp{\Wignerve[\uve]}}=\smash{\esp{\Wignerv_0}}$ by construction, and one gets:
\begin{multline}\label{eq:RTE1}
\partial_t\esp{\speciv_\jeig}+\{\eigl_\jeig,\esp{\speciv_\jeig}\}+\vl_\jeig\esp{\speciv_\jeig} + \esp{\speciv_\jeig}\adj{\vl}_\jeig \\
+ \nv_\jeig\esp{\speciv_\jeig} - \esp{\speciv_\jeig}\nv_\jeig =\esp{{\mathcal I}_1}-\esp{{\mathcal I}_2}\,,
\end{multline}
where:
\begin{equation}\label{eq:I1}
\begin{split}
{\mathcal I}_1 &=\ci\wg\adj{\eigc}_\jeig(\kv)\Wignerv_\half(\yv,\kv)\adj{\Kper}(\yv)\eigc_\jeig(\kv) \\
&=\sum_{\keig}\iint\iexp^{\ci\yv\cdot(\vp-\vq)}\frac{\eigl_\keig^2(\kv-\vp)\kper_{\jeig\keig}(\kv,\vp,\kv-\vp)\speciv_\keig(\kv-\vp)\kper_{\keig\jeig}(\kv-\vp,-\vq,\kv)}{\ci(\eigl_\jeig(\kv)-\eigl_\keig(\kv-\vp))+\theta}\frac{\dd\vp\dd\vq}{(2\pi)^6} \\
&\quad-\eigl_\jeig^2(\kv)\speciv_\jeig(\kv)\adj{\left(\sum_{\keig}\iint\iexp^{-\ci\yv\cdot(\vp-\vq)}\frac{\kper_{\jeig\keig}(\kv,\vq,\kv-\vp)\kper_{\keig\jeig}(\kv-\vp,-\vp,\kv)}{\ci(\eigl_\keig(\kv-\vp)-\eigl_\jeig(\kv))+\theta}\frac{\dd\vp\dd\vq}{(2\pi)^6}\right)}\,;
\end{split}
\end{equation}
and:
\begin{equation}\label{eq:I2}
{\scriptsize
\begin{split}
{\mathcal I}_2 &=\ci\wg \adj{\eigc}_\jeig(\kv)\left(\iint \iexp^{\ci\yv\cdot(\vp+\vq)} \TF{\Kper} (\vp) \TF{\Wignerv}_\half(\vq,\kv-\vp) \frac{\dd\vp\dd\vq}{(2\pi)^6}\right)\eigc_\jeig(\kv) \\
&=\sum_{\gamma,\keig}\left(\iint \iexp^{\ci\yv\cdot(\vp+\vq)} \frac{\eigl_\keig^2(\kv-\vp-\vq)\kper_{\jeig\gamma}(\kv,\vp,\kv-\vp)\kper_{\gamma\keig}(\kv-\vp,\vq,\kv-\vp-\vq)\speciv_\keig(\kv-\vp-\vq)\adj{\eigb}_\keig(\kv-\vp-\vq)}{\ci(\eigl_\gamma(\kv-\vp)-\eigl_\keig(\kv-\vp-\vq))+\theta} \frac{\dd\vp\dd\vq}{(2\pi)^6}\right)\eigc_\jeig(\kv) \\
&\quad - \sum_{\gamma,\keig}\left(\iint \iexp^{\ci\yv\cdot(\vp+\vq)} \frac{\eigl_\gamma^2(\kv-\vp)\kper_{\jeig\gamma}(\kv,\vp,\kv-\vp)\speciv_\gamma(\kv-\vp)\kper_{\gamma\keig}(\kv-\vp,\vq,\kv-\vp-\vq)\adj{\eigb}_\keig(\kv-\vp-\vq)}{\ci(\eigl_\gamma(\kv-\vp)-\eigl_\keig(\kv-\vp-\vq))+\theta} \frac{\dd\vp\dd\vq}{(2\pi)^6}\right)\eigc_\jeig(\kv)\,.
\end{split}}
\end{equation}
In the above relationships we have introduced:
\begin{displaymath}
\kper_{\jeig\keig}(\kv,\vp,\vq):=\adj{\eigc}_\jeig(\kv)\TF{\Kper}(\vp)\eigb_\keig(\vq)=\adj{\kper_{\keig\jeig}(\vq,-\vp,\kv)}\,,
\end{displaymath}
where the second equality stems from $\Kv_\iref\Kper=\adj{\Kper}\Kv_\iref$ by the Hermiticity of the optical response. By the change of variable $-\vq\to\vq$, the use of \eref{eq:DSP}, and then the change of variable $\kv-\vp\to\vp$, the average of the first sum in \eref{eq:I1} reads:
\begin{displaymath}
\begin{split}
\esp{{\mathcal I}_{11}} &=\sum_{\keig}\iint\iexp^{\ci\yv\cdot(\vp+\vq)}\frac{\eigl_\keig^2(\kv-\vp)\esp{\kper_{\jeig\keig}(\kv,\vp,\kv-\vp)\speciv_\keig(\kv-\vp)\kper_{\keig\jeig}(\kv-\vp,\vq,\kv)}}{\ci(\eigl_\jeig(\kv)-\eigl_\keig(\kv-\vp))+\theta}\frac{\dd\vp\dd\vq}{(2\pi)^6} \\
&=\sum_{\keig}\int\frac{\eigl_\keig^2(\kv-\vp)\TF{\coro}_{\jeig\keig}(\kv,\kv-\vp):\esp{\speciv_\keig(\kv-\vp)}}{\ci(\eigl_\jeig(\kv)-\eigl_\keig(\kv-\vp))+\theta}\dd\vp \\
&=\sum_{\keig}\int\frac{\eigl_\keig^2(\vp)\TF{\coro}_{\jeig\keig}(\kv,\vp):\esp{\speciv_\keig(\vp)}}{\ci(\eigl_\jeig(\kv)-\eigl_\keig(\vp))+\theta}\dd\vp \\
\end{split}
\end{displaymath}
where the linear operator $\smash{\TF{\coro}_{\jeig\keig}}$ is:
\begin{equation*}
[\TF{\coro}_{\jeig\keig}(\kv,\vp)]_{aa'bb'} :=\cjg{\eigc_\jeig^a(\kv)}\otimes\eigb_\keig^b(\vp):\TF{\coro}(\kv-\vp):\cjg{\eigc_\keig^{b'}(\vp)}\otimes\eigb_\jeig^{a'}(\kv)\,,
\end{equation*}
for $1\leq a,a'\leq A$ and $1\leq b,b'\leq B$, $A$ and $B$ being the orders of multiplicity of the modes $\jeig$ and $\keig$, respectively.\footnote{This result should be compared with \cite[Eq. (42)]{FAN07b} or \cite[Eq. (42)]{FAN07c} where no coupling between modes occurs and this operator rather reads:
\begin{displaymath}[\TF{\coro}_\jeig(\kv,\vp)]_{aa'bb'} :=\cjg{\eigc_\jeig^a(\kv)}\otimes\eigb_\jeig^{b}(\vp):\TF{\coro}(\kv-\vp):\eigc_\jeig^{a'}(\kv)\otimes\cjg{\eigb_\jeig^{b'}(\vp)}\,.
\end{displaymath}}
Here we have used a mixing assumption by which expectations over products of $\smash{\TF{\Kper}}$ and over $\smash{\speciv_\keig}$ are independent and get decoupled, because both quantities vary on different scales. Now using the changes of variable $-\vp\to\vp$, and then $\kv+\vp\to\vp$, the average of the second sum in \eref{eq:I1} reads:
\begin{equation}\label{eq:tscat1}
\begin{split}
\esp{{\mathcal I}_{12}} &=\eigl_\jeig^2(\kv)\sum_{\keig}\iint\iexp^{\ci\yv\cdot(\vq-\vp)}\frac{\esp{\kper_{\jeig\keig}(\kv,\vq,\kv-\vp)\kper_{\keig\jeig}(\kv-\vp,-\vp,\kv)}}{\ci(\eigl_\keig(\kv-\vp)-\eigl_\jeig(\kv))+\theta}\frac{\dd\vp\dd\vq}{(2\pi)^6} \\
&=\eigl_\jeig^2(\kv)\sum_{\keig}\iint\iexp^{\ci\yv\cdot(\vq+\vp)}\frac{\esp{\kper_{\jeig\keig}(\kv,\vq,\kv+\vp)\kper_{\keig\jeig}(\kv+\vp,\vp,\kv)}}{\ci(\eigl_\keig(\kv+\vp)-\eigl_\jeig(\kv))+\theta}\frac{\dd\vp\dd\vq}{(2\pi)^6} \\
&=\sum_{\keig}\int\frac{\eigl_\jeig^2(\kv)\TF{\coro}_{\jeig\keig}(\kv,\kv+\vp):\Id_B}{\ci(\eigl_\keig(\kv+\vp)-\eigl_\jeig(\kv))+\theta}\dd\vp \\
&=\sum_{\keig}\int\frac{\eigl_\jeig^2(\kv)\TF{\coro}_{\jeig\keig}(\kv,\vp):\Id_B}{\ci(\eigl_\keig(\vp)-\eigl_\jeig(\kv))+\theta}\dd\vp\,.
\end{split}
\end{equation}
Besides, taking the average of \eref{eq:I2}, one has with \eref{eq:DSP}:
\begin{displaymath}
\begin{split}
\esp{{\mathcal I}_2} &=\left(\sum_{\gamma}\int\frac{\eigl_\jeig^2(\kv)\TF{\coro}_{\jeig\gamma}(\kv,\kv-\vp):\Id_G}{\ci(\eigl_\gamma(\kv-\vp)-\eigl_\jeig(\kv))+\theta} \dd\vp\right)\esp{\speciv_\jeig(\kv)} \\
&\quad\quad - \sum_{\gamma}\int \frac{\eigl_\gamma^2(\kv-\vp)\TF{\coro}_{\jeig\gamma}(\kv,\kv-\vp):\esp{\speciv_\gamma(\kv-\vp)}}{\ci(\eigl_\gamma(\kv-\vp)-\eigl_\jeig(\kv))+\theta} \dd\vp \\
&=\left(\sum_{\gamma}\int\frac{\eigl_\jeig^2(\kv)\TF{\coro}_{\jeig\gamma}(\kv,\vp):\Id_G}{\ci(\eigl_\gamma(\vp)-\eigl_\jeig(\kv))+\theta} \dd\vp\right)\esp{\speciv_\jeig(\kv)} \\
&\quad\quad - \sum_{\gamma}\int \frac{\eigl_\gamma^2(\vp)\TF{\coro}_{\jeig\gamma}(\kv,\vp):\esp{\speciv_\gamma(\vp)}}{\ci(\eigl_\gamma(\vp)-\eigl_\jeig(\kv))+\theta} \dd\vp\,,
\end{split}
\end{displaymath}
where $G$ stands for the order of multiplicity of the mode $\gamma$.

The last step is to let $\theta\to 0$, which is done thank to the (formal) identity $\smash{\frac{1}{\ci x+\theta}}\to\smash{\frac{1}{\ci x}}+\hat{\theta}\pi\delta(x)$, for $\smash{\hat{\theta}}$ being the sign of $\theta$. This yields:
\begin{multline*}
\esp{{\mathcal I}_1}-\esp{{\mathcal I}_2}=\sum_{\keig}\int 2\pi\hat{\theta}\delta(\eigl_\jeig(\kv)-\eigl_\keig(\vp))\eigl_\keig^2(\vp)\TF{\coro}_{\jeig\keig}(\kv,\vp):\esp{\speciv_\keig(\vp)}\dd\vp \\
-\esp{\speciv_\jeig(\kv)}\adj{\left[\sum_{\keig}\int\left(\frac{1}{\ci(\eigl_\keig(\vp)-\eigl_\jeig(\kv))}+\pi\hat{\theta}\delta(\eigl_\keig(\vp)-\eigl_\jeig(\kv))\right)\eigl_\jeig^2(\kv)\TF{\coro}_{\jeig\keig}(\kv,\vp):\Id_B\dd\vp\right]} \\
-\left[\sum_{\keig}\int\left(\frac{1}{\ci(\eigl_\keig(\vp)-\eigl_\jeig(\kv))}+\pi\hat{\theta}\delta(\eigl_\keig(\vp)-\eigl_\jeig(\kv))\right)\eigl_\jeig^2(\kv)\TF{\coro}_{\jeig\keig}(\kv,\vp):\Id_B\dd\vp\right]\esp{\speciv_\jeig(\kv)}\,.
\end{multline*}
Consider $\hat{\theta}=1$ (for causality purposes) and let:
\begin{equation}\label{eq:dscat}
\dscat_{\jeig\keig}(\kv,\vp) = 2\pi\eigl_\jeig(\kv)\eigl_\keig(\vp)\TF{\coro}_{\jeig\keig}(\kv,\vp)\,,
\end{equation}
and:
\begin{equation}\label{eq:tscat}
\tscat_\jeig(\kv)=\demi\sum_{\keig}\int\delta(\eigl_\keig(\vp)-\eigl_\jeig(\kv))\dscat_{\jeig\keig}(\kv,\vp):\Id_B\dd\vp -\frac{\ci}{2\pi}\sum_\keig\oint\frac{\dscat_{\jeig\keig}(\kv,\vp):\Id_B}{\eigl_\keig(\vp)-\eigl_\jeig(\kv)}\dd\vp\,,
\end{equation}
where the second integral holds as a Cauchy principal value; then \eref{eq:RTE1} finally reads as the radiative transfer equations:
\begin{multline}\label{eq:RTEaverage}
\partial_t\esp{\speciv_\jeig}+\{\eigl_\jeig,\esp{\speciv_\jeig}\}+\vl_\jeig\esp{\speciv_\jeig} +\esp{\speciv_\jeig}\adj{\vl}_\jeig +\nv_\jeig\esp{\speciv_\jeig} - \esp{\speciv_\jeig}\nv_\jeig \\
=\sum_{\keig}\int\delta(\eigl_\keig(\vp)-\eigl_\jeig(\kv))\dscat_{\jeig\keig}(\kv,\vp):\esp{\speciv_\keig(\vp)}\dd\vp \\
 - \tscat_\jeig(\kv)\esp{\speciv_\jeig(\kv)} - \esp{\speciv_\jeig(\kv)}\adj{\tscat_\jeig(\kv)} \,,
\end{multline}
possibly coupling all modes of propagation while keeping the frequency constant: $\eigl_\keig(\vp)=\eigl_\jeig(\kv)$ in the scattering processes. This result extends \cite[Eq. (4.32)]{RYZ96} for isotropic dielectric media to the case of general bianisotropic dielectric media.\footnote{The sole difference lies in the sign of the imaginary part of $\tscat_\jeig$ and the $1/\pi$ factor, to be compared with \cite[Eq. (4.34)]{RYZ96}.} It revisits the derivation in \cite{FAN07b,FAN07c} by considering the time-dependent case with damping effects, and possible coupling between the modes. Moreover, it uses Wigner transforms and pseudo-differential calculus in the same standard quantization rather than mixing it with the Weyl quantization. Matrix-valued radiative transfer equations of the form \eqref{eq:RTEaverage} can be solved numerically by the Monte-Carlo method, which is based on their probabilistic representation in terms of a jump Markov process as outlined in \cite{PAP00}. It extends ray tracing to scattering media.

\begin{remark}
In \cite{BUT15} the derivation of a radiative transfer equation of the form of \eref{eq:RTEaverage} is proved rigorously for the scalar wave equation with a random speed of sound. Here it is shown that the average Wigner transform of the wave field converges (weakly) to the solution of a linear Boltzmann equation similar to \eqref{eq:RTEaverage}, provided that the initial conditions satisfy some \emph{adhoc} boundedness and tightness conditions and that the random fluctuations fulfill some regularity assumptions as well. A self-averaging property is also demonstrated, by which the deviations of the Wigner transform from its average are shown to vanish in the high-frequency limit. This means that \eref{eq:RTEaverage} would also hold for $\speciv_\jeig$ and not only $\esp{\speciv_\jeig}$, a result that is implicitly acknowledged in the statement of \eref{eq:RTE}. The proofs in \cite{BUT15} are however quite involved and very technical so that their extension to the present case is much beyond the scope of this research. Radiative transfer equations can also be rigorously derived for some particular configurations as in \cite{ERD00,LUK07}. 
\end{remark}

\begin{example}
For the Lorentz model with damping (\ref{eq:Lorentz}) the scattering cross-sections \eqref{eq:dscat} are the ones derived in \cite{RYZ96} for isotropic media with eigenvalues $\eigl_\pm(\kv)=\pm\cel_\iref\norm{\kv}$, each of multiplicity two, for the propagating modes. Couplings in \eref{eq:RTEaverage} occur only for those eigenvalues with the same wave numbers $\norm{\vp}=\norm{\kv}$ and signs. The corresponding right eigenvectors $\smash{\eigb_\pm^a(\hkv)}$ are given by \eref{eq:eigv-Lorentz} and the left eigenvectors are for $a=1,2$:
\begin{displaymath}
\eigc_\pm^a(\hkv)=\begin{pmatrix} \displaystyle\sqrt{\frac{\permitj_\iref}{2}}\ev_a(\hkv)\\ \pm\displaystyle\sqrt{\frac{\permeaj_\iref}{2}}\hkv\times\ev_a(\hkv) \end{pmatrix}\,.
\end{displaymath}
Besides, the dimensionless random fluctuations $\Kper(\yv)$ read:
\begin{displaymath}
\Kper(\yv)=\begin{bmatrix} \permitj_1(\yv)\Id & \bzero \\ \bzero & \permeaj_1(\yv)\Id \end{bmatrix}\,,
\end{displaymath}
where $\smash{(\permitj_1(\yv),\,\yv\in\Rset^3)}$ is the dimensionless random fluctuation of the permittivity $\smash{\permitj_\iref}$, and $\smash{(\permeaj_1(\yv),\,\yv\in\Rset^3)}$ is the dimensionless random fluctuation of the permeability $\smash{\permeaj_\iref}$. Both are real-valued, second-order, mean-square homogeneous random processes  with zero means. Their power spectral density functions are denoted by $\smash{\TF{\coroij}_\permitj(\kv)}$ and $\smash{\TF{\coroij}_\permeaj(\kv)}$, respectively, and their cross-power spectral density function is denoted by $\smash{\TF{\coroij}_{\permitj\permeaj}(\kv)}=\smash{\TF{\coroij}_{\permeaj\permitj}(-\kv)}$.
Introducing the $2\times 2$ matrices $\smash{\vT(\hkv,\hvp)}$ and $\smash{\vX(\hkv,\hvp)}$ such that:
\begin{displaymath}
T_{ab}(\hkv,\hvp)=\ev_a(\hkv)\cdot\ev_b(\hvp)\,,\quad X_{ab}(\hkv,\hvp)=(\hkv\times\ev_a(\hkv))\cdot(\hvp\times\ev_b(\hvp))\,,
\end{displaymath}
we have $\smash{\adj{\vT(\hkv,\hvp)}}=\smash{\vT(\hvp,\hkv)}$, $\smash{\adj{\vX(\hkv,\hvp)}}=\smash{\vX(\hvp,\hkv)}$, and $\smash{\vT(\hkv,\hvp)\vX(\hvp,\hkv)}=\smash{(\hkv\cdot\hvp)}\Id$. The differential scattering kernels in \eref{eq:RTEaverage} then read:
\begin{displaymath}
\begin{split}
\dscat_{\jeig\jeig}(\kv,\vp):\speciv_\jeig(\vp)=\frac{\pi}{2}\cel_\iref^2\norm{\kv}\norm{\vp}\Big[ & \TF{\coroij}_\permitj(\kv-\vp)\vT(\hkv,\hvp)\speciv_\jeig(\vp)\vT(\hvp,\hkv) \\
& +\TF{\coroij}_\permeaj(\kv-\vp)\vX(\hkv,\hvp)\speciv_\jeig(\vp)\vX(\hvp,\hkv) \\
& + \TF{\coroij}_{\permitj\permeaj}(\kv-\vp)\vT(\hkv,\hvp)\speciv_\jeig(\vp)\vX(\hvp,\hkv) \\
& +  \TF{\coroij}_{\permeaj\permitj}(\kv-\vp)\vX(\hkv,\hvp)\speciv_\jeig(\vp)\vT(\hvp,\hkv)  \Big]\,,
\end{split}
\end{displaymath}
and $\dscat_{\jeig\keig}:\speciv_\keig=\bzero$ whenever $\keig\neq\jeig$, $\jeig,\keig\in\{+,-\}$. If in addition the random fluctuations $\smash{\permitj_1}$ and $\smash{\permeaj_1}$ are isotropic, which means that $\smash{\TF{\coroij}_\permitj(\kv)}$, $\smash{\TF{\coroij}_\permeaj(\kv)}$, and $\smash{\TF{\coroij}_{\permitj\permeaj}(\kv)}$ depend on $\norm{\kv}$ solely, it can be shown that $\smash{\int_{S^2}\dscat_{\jeig\jeig}(\kv,\vp):\Id_A\dd\Omega(\hvp)}$ is proportional to the identity matrix. Then the total scattering cross-sections \eqref{eq:tscat} are diagonal and real, and the formula of \cite[Eq. (4.46)]{RYZ96} is recovered; see \aref{sec:tdscat}.
\end{example}

\begin{example}
A chiral medium has optical response:
\begin{displaymath}
\Kv_\iref=\begin{bmatrix} \permitj_\iref\Id & \ci\chi\Id \\ -\ci\chi\Id & \permeaj_\iref\Id \end{bmatrix}
\end{displaymath}
where $\chi\in\Rset$ is the magnetoelectric constant and $\kappa=\smash{\cel_\iref\chi}$ is the chirality parameter, which is such that $\abs{\kappa}<1$ to preserve positivity of $\smash{\Kv_\iref}$. We consider biisotropic perturbations as in \cite{FAN07b}:
 \begin{displaymath}
\Kper(\yv)=\begin{bmatrix} a(\yv)\Id & \ci\impj_\iref b(\yv)\Id \\ \frac{b(\yv)}{\ci\impj_\iref} \Id & a(\yv)\Id \end{bmatrix}\,,
\end{displaymath}
where $\smash{\impj_\iref}=\smash{\sqrt{\permeaj_\iref/\permitj_\iref}}$ is the impedance, and $\smash{(a(\yv),\,\yv\in\Rset^3)}$ and $\smash{(b(\yv),\,\yv\in\Rset^3)}$ are real-valued, second-order, mean-square homogeneous random processes with zero means. Their power spectral density functions are denoted by $\smash{\TF{\coroij}_a(\kv)}$ and $\smash{\TF{\coroij}_b(\kv)}$, respectively, and their cross-power spectral density function is denoted by $\smash{\TF{\coroij}_{ab}(\kv)}=\smash{\TF{\coroij}_{ba}(-\kv)}$. 
Then $\smash{\vL_\iref}$ has four non-zero simple eigenvalues:
\begin{displaymath}
\eigl_1(\kv)=\frac{\cel_\iref\norm{\kv}}{1+\kappa}\,,\quad\eigl_2(\kv)=-\frac{\cel_\iref\norm{\kv}}{1+\kappa}\,,\quad\eigl_3(\kv)=\frac{\cel_\iref\norm{\kv}}{1-\kappa}\,,\quad\eigl_4(\kv)=-\frac{\cel_\iref\norm{\kv}}{1-\kappa}\,,
\end{displaymath}
associated to the right eigenvectors:\footnote{Note that for $\kappa=0$, the eigenvectors \eqref{eq:eigv-Lorentz} are recovered taking the summation and difference of $\eigb_1$ and $\eigb_3$ on one hand, yielding $\eigb_+^1$ and $\eigb_+^2$, and the summation and difference of $\eigb_2$ and $\eigb_4$ on the other hand, yielding $\eigb_-^1$ and $\eigb_-^2$.}
\begin{displaymath}
\eigb_1(\hkv)=\frac{1}{\sqrt{1+\kappa}}\begin{pmatrix} \displaystyle \frac{\ev_1'(\hkv)}{\sqrt{2\permitj_\iref}} \\ \displaystyle\frac{\hkv\times\ev_1'(\hkv)}{\sqrt{2\permeaj_\iref}} \end{pmatrix}\,,\quad\eigb_2(\hkv)=\frac{1}{\sqrt{1+\kappa}}\begin{pmatrix} \displaystyle \frac{\ev_2'(\hkv)}{\sqrt{2\permitj_\iref}} \\ \displaystyle-\frac{\hkv\times\ev_2'(\hkv)}{\sqrt{2\permeaj_\iref}} \end{pmatrix}\,,
\end{displaymath}
and:
\begin{displaymath}
\eigb_3(\hkv)=\frac{1}{\sqrt{1-\kappa}}\begin{pmatrix} \displaystyle  \frac{\ev_2'(\hkv)}{\sqrt{2\permitj_\iref}} \\ \displaystyle\frac{\hkv\times\ev_2'(\hkv)}{\sqrt{2\permeaj_\iref}} \end{pmatrix}\,,\quad\eigb_4(\hkv)=\frac{1}{\sqrt{1-\kappa}}\begin{pmatrix} \displaystyle \frac{\ev_1'(\hkv)}{\sqrt{2\permitj_\iref}} \\ \displaystyle-\frac{\hkv\times\ev_1'(\hkv)}{\sqrt{2\permeaj_\iref}} \end{pmatrix}\,,
\end{displaymath}
letting $\ev_1'(\hkv)=\frac{1}{\sqrt{2}}(\ci\ev_1(\hkv)-\ev_2(\hkv))$ and $\ev_2'(\hkv)=\frac{1}{\sqrt{2}}(\ci\ev_1(\hkv)+\ev_2(\hkv))$. The left eigenvectors are then:
\begin{displaymath}
\eigc_1(\hkv)=\sqrt{1+\kappa}\begin{pmatrix} \displaystyle \sqrt{\frac{\permitj_\iref}{2}}\ev_1'(\hkv) \\ \displaystyle\sqrt{\frac{\permeaj_\iref}{2}}\hkv\times\ev_1'(\hkv) \end{pmatrix}\,,\quad\eigc_2(\hkv)=\sqrt{1+\kappa}\begin{pmatrix} \displaystyle\sqrt{\frac{\permitj_\iref}{2}}\ev_2'(\hkv) \\ \displaystyle-\sqrt{\frac{\permeaj_\iref}{2}}\hkv\times\ev_2'(\hkv) \end{pmatrix}\,,
\end{displaymath}
and:
\begin{displaymath}
\eigc_3(\hkv)=\sqrt{1-\kappa}\begin{pmatrix} \displaystyle \sqrt{\frac{\permitj_\iref}{2}}\ev_2'(\hkv) \\  \displaystyle \sqrt{\frac{\permeaj_\iref}{2}}\hkv\times\ev_2'(\hkv) \end{pmatrix}\,,\quad\eigc_4(\hkv)=\sqrt{1-\kappa}\begin{pmatrix} \displaystyle \sqrt{\frac{\permitj_\iref}{2}}\ev_1'(\hkv) \\ \displaystyle-\sqrt{\frac{\permeaj_\iref}{2}}\hkv\times\ev_1'(\hkv) \end{pmatrix}\,.
\end{displaymath}
All scattering cross-sections are scalars and couplings in \eref{eq:RTEaverage} possibly occur for those eigenvalues with the same signs and $\norm{\vp}=\norm{\kv}$, or $(1\pm\kappa)\norm{\vp}=(1\mp\kappa)\norm{\kv}$. However one obtains $\dscatij_{13}(\kv,\vp)=\dscatij_{24}(\kv,\vp)=0$ and only self-couplings occur with scattering cross-sections:
\begin{displaymath}
\begin{split}
\dscatij_{11}(\kv,\vp) &=\frac{2\pi\cel_\iref^2\norm{\kv}\abs{\vp}}{(1+\kappa)^2}\abs{\cjg{\ev_1'(\hkv)}\cdot\ev_1'(\hvp)}^2\left(\TF{\coroij}_a(\kv-\vp)+\TF{\coroij}_b(\kv-\vp)+2\TF{\coroij}_{ab}^s(\kv-\vp)\right)\,, \\
\dscatij_{22}(\kv,\vp) &=\frac{2\pi\cel_\iref^2\norm{\kv}\abs{\vp}}{(1+\kappa)^2}\abs{\cjg{\ev_2'(\hkv)}\cdot\ev_2'(\hvp)}^2\left(\TF{\coroij}_a(\kv-\vp)+\TF{\coroij}_b(\kv-\vp)+2\TF{\coroij}_{ab}^s(\kv-\vp)\right)\,, \\
\dscatij_{33}(\kv,\vp) &=\frac{2\pi\cel_\iref^2\norm{\kv}\abs{\vp}}{(1-\kappa)^2}\abs{\cjg{\ev_2'(\hkv)}\cdot\ev_2'(\hvp)}^2\left(\TF{\coroij}_a(\kv-\vp)+\TF{\coroij}_b(\kv-\vp)-2\TF{\coroij}_{ab}^s(\kv-\vp)\right)\,, \\
\dscatij_{44}(\kv,\vp) &=\frac{2\pi\cel_\iref^2\norm{\kv}\abs{\vp}}{(1-\kappa)^2}\abs{\cjg{\ev_1'(\hkv)}\cdot\ev_1'(\hvp)}^2\left(\TF{\coroij}_a(\kv-\vp)+\TF{\coroij}_b(\kv-\vp)-2\TF{\coroij}_{ab}^s(\kv-\vp)\right)\,,
\end{split}
\end{displaymath}
for $\smash{\TF{\coroij}_{ab}^s(\kv)}=\smash{\demi(\TF{\coroij}_{ab}(\kv)+\TF{\coroij}_{ba}(\kv))}$. The total scattering cross-sections for statistically isotropic perturbations of the optical response are then:
\begin{multline*}
\tscati_1(\norm{\kv})=\tscati_2(\norm{\kv})=\frac{\pi^2\cel_\iref\norm{\kv}^4}{1+\kappa}\int_{-1}^1(1+\Theta)^2\Big[\TF{\coroij}_a\left(\norm{\kv}\sqrt{2(1-\Theta)}\right)+\TF{\coroij}_b\left(\norm{\kv}\sqrt{2(1-\Theta)}\right) \\
+2\TF{\coroij}_{ab}\left(\norm{\kv}\sqrt{2(1-\Theta)}\right)\Big]\dd\Theta\,,
\end{multline*}
since $\TF{\coroij}_{ab}(\norm{\kv})=\TF{\coroij}_{ba}(\norm{\kv})$ in the isotropic case, and:
\begin{multline*}
\tscati_3(\norm{\kv})=\tscati_4(\norm{\kv})=\frac{\pi^2\cel_\iref\norm{\kv}^4}{1-\kappa}\int_{-1}^1(1+\Theta)^2\Big[\TF{\coroij}_a\left(\norm{\kv}\sqrt{2(1-\Theta)}\right)+\TF{\coroij}_b\left(\norm{\kv}\sqrt{2(1-\Theta)}\right) \\
-2\TF{\coroij}_{ab}\left(\norm{\kv}\sqrt{2(1-\Theta)}\right)\Big]\dd\Theta\,.
\end{multline*}
\end{example}

\section{Summary and outlook}\label{sec:CL}

We have derived a system of coupled radiative transfer equations to describe the propagation of high-frequency electromagnetic waves in randomly fluctuating bianisotropic dielectric media including dispersive and dissipative effects. The fluctuations are weak and vary at the same length scales as the typical wavelength. We have used a Wigner functional approach and its interpretation in terms of semiclassical pseudo-differential operators to derive the transfer equations from a multiscale analysis. Radiative transfer equations have a geometrical interpretation in terms of bicharacteristic rays that is well suited for their numerical integration by ray methods and ray tracing solvers. Our results extend this approach to heterogeneous {\em bianisotropic} media or plasmas beyond the classical applications of ray tracing in homogeneous or piecewise homogeneous {\em isotropic} media. In future works one aims to derive the diffusion limit of the radiative transfer model, which describes the evolution of the energy density in physical space solely after the waves have travelled several scattering mean free paths and have lost the memory of their angular distribution at earlier times. The issue of deriving boundary conditions adapted to the Wigner distribution from the boundary conditions applied to the electromagnetic fields is also worth considering, for applications in radar imaging or remote sensing with scenes exhibiting sharp interfaces for example.

\section*{Acknowledgement}

We thank Thomas Lepetit at ONERA for valuable discussions.

\appendix

\section{Total scattering cross-sections for the Lorentz model}\label{sec:tdscat}

We show here that $\smash{\int_{S^2}\dscat_{\jeig\jeig}(\kv,\vp):\Id_A\dd\Omega(\hvp)}$ is real and proportional to the identity matrix for isotropic fluctuations of the permittivity and permeability. Then we derive the total scattering cross-sections \eqref{eq:tscat}. We first obtain by direct computation:
\begin{displaymath}
\vT(\hkv,\hvp)\vT(\hvp,\hkv)=\begin{bmatrix} 1-(\hvp\cdot\ev_1(\hkv))^2 & -(\hvp\cdot\ev_1(\hkv))(\hvp\cdot\ev_2(\hkv)) \\ -(\hvp\cdot\ev_1(\hkv))(\hvp\cdot\ev_2(\hkv)) & 1-(\hvp\cdot\ev_2(\hkv))^2 \end{bmatrix}\,,
\end{displaymath}
and:
\begin{displaymath}
\vX(\hkv,\hvp)\vX(\hvp,\hkv)=\begin{bmatrix} 1-(\hvp\cdot\ev_2(\hkv))^2 & (\hvp\cdot\ev_1(\hkv))(\hvp\cdot\ev_2(\hkv)) \\ (\hvp\cdot\ev_1(\hkv))(\hvp\cdot\ev_2(\hkv)) & 1-(\hvp\cdot\ev_1(\hkv))^2 \end{bmatrix}\,.
\end{displaymath}
We then write $\vp=\norm{\vp}(\sin\theta\cos\phi,\sin\theta\sin\phi,\cos\theta)^\itr$ in the right-handed orthonormal frame $\smash{(\ev_1(\hkv),\ev_2(\hkv),\hkv)}$, such that $\smash{\hkv\cdot\hvp}=\cos\theta$ and $\dd\Omega(\hvp)=\sin\theta\dd\theta\dd\phi=-\dd(\cos\theta)\dd\phi$ for $\theta\in[0,\pi]$ and $\phi\in[0,2\pi]$. Consequently one obtains:
\begin{displaymath}
{\scriptsize
\begin{split}
\int_{S^2} \TF{\coroij}_\permitj(\kv-\vp)\vT(\hkv,\hvp)\vT(\hvp,\hkv)\dd\Omega(\hvp) &=\int_0^\pi\int_0^{2\pi}\TF{\coroij}_\permitj(\kv-\vp)\begin{bmatrix} 1-\sin^2\theta\cos^2\phi & -\demi\sin^2\theta\sin2\phi \\ -\demi\sin^2\theta\sin2\phi  & 1-\sin^2\theta\sin^2\phi \end{bmatrix}\sin\theta\dd\theta\dd\phi \\
& =\pi\left(\int_0^\pi\TF{\coroij}_\permitj\left(\sqrt{\norm{\kv}^2+\norm{\vp}^2-2\norm{\kv}\norm{\vp}\cos\theta}\right) (1+\cos^2\theta)\sin\theta\dd\theta\right)\Id
\end{split}}
\end{displaymath}
provided that $\kv\mapsto\smash{\TF{\coroij}_\permitj(\kv)}$ depends on $\norm{\kv}$ solely, or:
\begin{multline*}
\int_{S^2} \TF{\coroij}_\permitj(\kv-\vp)\vT(\hkv,\hvp)\vT(\hvp,\hkv)\dd\Omega(\hvp) \\
=\pi\left(\int_{-1}^1\TF{\coroij}_\permitj\left(\sqrt{\norm{\kv}^2+\norm{\vp}^2-2\norm{\kv}\norm{\vp}\Theta}\right) (1+\Theta^2)\dd\Theta\right)\Id\,.
\end{multline*}
Likewise, one obtains:
\begin{multline*}
\int_{S^2} \TF{\coroij}_\permeaj(\kv-\vp)\vX(\hkv,\hvp)\vX(\hvp,\hkv)\dd\Omega(\hvp) \\
=\pi\left(\int_{-1}^1\TF{\coroij}_\permeaj\left(\sqrt{\norm{\kv}^2+\norm{\vp}^2-2\norm{\kv}\norm{\vp}\Theta}\right) (1+\Theta^2)\dd\Theta\right)\Id
\end{multline*}
provided that $\kv\mapsto\smash{\TF{\coroij}_\permeaj(\kv)}$ also depends on $\norm{\kv}$ solely. Lastly:
\begin{multline*}
\int_{S^2} \left(\TF{\coroij}_{\permitj\permeaj}(\kv-\vp)\vT(\hkv,\hvp)\vX(\hvp,\hkv)+\TF{\coroij}_{\permitj\permeaj}(\vp-\kv)\vX(\hkv,\hvp)\vT(\hvp,\hkv)\right)\dd\Omega(\hvp) \\
=4\pi\left(\int_{-1}^1\TF{\coroij}_{\permitj\permeaj}\left(\sqrt{\norm{\kv}^2+\norm{\vp}^2-2\norm{\kv}\norm{\vp}\Theta}\right)\Theta\dd\Theta\right)\Id
\end{multline*}
provided that $\kv\mapsto\smash{\TF{\coroij}_{\permitj\permeaj}(\kv)}$ depends on $\norm{\kv}$ solely. Now if the power spectral density functions $\smash{\TF{\coroij}_\permitj}$, $\smash{\TF{\coroij}_\permeaj}$, and $\smash{\TF{\coroij}_{\permitj\permeaj}}$ are real, we see that the second integral (the Cauchy principal value) in \eref{eq:tscat} is real, proportional to the identity, and thus its contribution to the radiative transfer equation \eqref{eq:RTEaverage} vanishes. These conditions are equivalent to:
\begin{displaymath}
\TF{\coroij}_\permitj(\kv)=\TF{\coroij}_\permitj(-\kv)\,,\quad\TF{\coroij}_\permeaj(\kv)=\TF{\coroij}_\permeaj(-\kv)\,,\quad\TF{\coroij}_{\permitj\permeaj}(\kv)=\TF{\coroij}_{\permitj\permeaj}(-\kv)\,,
\end{displaymath}
which actually hold when these functions depend on $\norm{\kv}$ solely. They also hold when the covariance functions $\yv\mapsto\smash{\coroij_\permitj(\yv)}$, $\yv\mapsto\smash{\coroij_\permeaj(\yv)}$, and $\yv\mapsto\smash{\coroij_{\permitj\permeaj}(\yv)}$ (see \eref{eq:covariance}) are even.

Gathering all these results, we conclude that the total scattering cross-sections $\tscat_\jeig$ of \eref{eq:tscat} are:
\begin{displaymath}
\begin{split}
\tscat_\jeig(\norm{\kg}) &=\demi\int_{\Rset^3}\delta(\eigl_\jeig(\vp)-\eigl_\jeig(\kv))\dscat_{\jeig\jeig}(\kv,\vp):\Id_A\dd\vp \\
&=\demi\int_0^{+\infty}\delta(\cel_\iref\norm{\vp}-\cel_\iref\norm{\kv})\left(\int_{S^2}\dscat_{\jeig\jeig}(\kv,\vp):\Id_A\dd\Omega(\hvp)\right)\norm{\vp}^2\dd\norm{\vp} \\
&=\frac{\pi^2\cel_\iref\norm{\kv}^4}{4}\Bigg(\int_{-1}^1\Big[(1+\Theta^2)\left(\TF{\coroij}_\permitj\left(\norm{\kv}\sqrt{2(1-\Theta)}\right)+\TF{\coroij}_\permeaj\left(\norm{\kv}\sqrt{2(1-\Theta)}\right)\right) \\
&\quad\quad\quad\quad\quad\quad\quad\quad + 4\Theta\TF{\coroij}_{\permitj\permeaj}\left(\norm{\kv}\sqrt{2(1-\Theta)}\right)\Big]\dd\Theta\Bigg)\Id_A\,,
\end{split}
\end{displaymath}
ignoring the contribution from the Cauchy principal value. Indeed the latter reads:
\begin{displaymath}
\oint\frac{\dscat_{\jeig\jeig}(\kv,\vp):\Id_A}{\eigl_\jeig(\vp)-\eigl_\jeig(\kv)}\dd\vp =\lim_{\varepsilon\to 0}\int_\varepsilon^{+\infty}\dd r\int_{\abs{\norm{\vp}-\norm{\kv}}=r}\frac{\dscat_{\jeig\jeig}(\kv,\vp):\Id_A}{\cel_\iref(\norm{\vp}-\norm{\kv})}\norm{\vp}^2\dd\Omega(\hvp)
\end{displaymath}
that is, integrals on spheres which are real and proportional to the identity. Once multiplied by $\ci$, they do not contribute anymore to the radiative transfer equation  \eqref{eq:RTEaverage} as already stated above.
\end{document}